\documentclass[useAMS,usenatbib,usegraphicx]{mn2e}
\include{epsf}
\newcommand\aj{{AJ}}%
\newcommand\araa{{ARA\&A}}%
\newcommand\apj{{ApJ}}%
\newcommand\apjl{{ApJ}}%
\newcommand\apjs{{ApJS}}%
\newcommand\aap{{A\&A}}%
\newcommand\mnras{{MNRAS}}%
\newcommand\na{{New A}}%
\newcommand\pasj{{PASJ}}%
\title[Calibrating an updated SPH scheme within {\tt GCD+}]
{Calibrating an updated SPH scheme within {\tt GCD+}}
\author[Kawata et~al.]
 {D.~Kawata$^{1}$\thanks{E-mail: dka@mssl.ucl.ac.uk},
T.~Okamoto$^{2}$, B.K.~Gibson$^{3,4}$, D.J.~Barnes$^{1}$, R.~Cen$^{5}$
\\
$^{1}$ Mullard Space Science Laboratory, University College London,
Holmbury St. Mary, Dorking, Surrey, RH5 6NT
\\
$^{2}$ Center for Computational Sciences, University of Tsukuba, 1-1-1, 
Tennodai, Tsukuba, Ibaraki, 305-8577, Japan
\\
$^{3}$ Jeremiah Horrocks Institute, 
University of Central Lancashire, Preston, PR1~2HE, UK
\\
$^{4}$ Monash Centre for Astrophysics, Monash University, VIC 3800, Australia
\\
$^{5}$ Princeton University Observatory, Princeton University, Princeton, 
NJ, 08544, USA
}
\date{Accepted .
      Received ;
      in original form }

\pagerange{\pageref{firstpage}--\pageref{lastpage}}
\pubyear{2009}

\begin{document}

\maketitle

\label{firstpage}

\begin{abstract}
We adapt a modern scheme of smoothed particle hydrodynamics (SPH) to our tree N-body/SPH galactic
chemodynamics code {\tt GCD+}.  The applied scheme includes implementations of the artificial viscosity switch and artificial thermal conductivity proposed by \citet{jm97}, \citet{rp07} and \citet{dp08}, to model discontinuities and Kelvin-Helmholtz instabilities more
accurately. We first present hydrodynamics test simulations and contrast the results to runs undertaken without artificial viscosity switch or thermal conduction. In addition, we also explore 
the different levels of smoothing by adopting larger or smaller smoothing lengths, i.e. a larger or smaller number of neighbour particles, $N_{\rm nb}$.
We demonstrate that the new version of {\tt GCD+} is capable of modelling Kelvin-Helmholtz instabilities to a similar level as the mesh code, {\tt Athena}. From the Gresho vortex, point-like explosion and self-similar collapse tests, we conclude that setting the smoothing length to keep the number of neighbour particles as high as $N_{\rm nb}\sim58$ is preferable to adopting smaller smoothing lengths. We present our optimised parameter sets from the hydrodynamics tests.
\end{abstract}

\begin{keywords}
hydrodynamics --- instabilities --- methods: N-body simulation
\end{keywords}

\section{Introduction}
\label{intro-sec}

Since it was introduced by \citet{ll77} and \citet{gm77},
the smoothed particle hydrodynamics (SPH) methodology has become a
regular tool for the numerical simulation of a wide range of
astronomical phenomena.  \citet{hk89} were the first to suggest that the
SPH approach would also prove invaluable in the simulation of galaxy
formation and evolution. Since then, a number of SPH codes have been
developed to simulate such systems, incorporating various physical
processes ranging from radiative cooling to star formation and
supernovae (SNe) feedback
\citep[e.g.][]{nk92,nw93,kwh96,sm95,myn99,clc98,dk99,sgv99,syw01,
ck04,gmw04,vs05,ssk06,msd08,onb08,sd08,sdk08,mbg10,vs10b,swp12}. 

Parallel to the development of such particle-based codes, grid- or
mesh-based approaches have been employed for modeling the formation and
evolution of galaxies \citep[e.g.][]{co92}.  Algorithmic enhancements to
a fixed grid approach, such as adaptive mesh refinement (AMR), has led to
a massive improvement in the capability of grid-based codes for
simulations which require a large dynamic range, including those of
galaxy formation \citep{rt02,ak03,tb08,gcs09,jcb09,scgb09}.  Code comparisons
between SPH and AMR
\citep{fwb99,aymg03,vkb05,onshn05,gcs09,tbm08,mmbtc09,hbg11,pfg12} demonstrate that
the competing approaches lead to generally consistent results.  That
said, comparing the results of hydrodynamics simulations of the
formation of a galaxy cluster, \citet{fwb99} claim that SPH codes lead
to lower entropy in the central region of the simulated cluster
\citep[see also][]{aymg03,vkb05,onshn05,dvbt05,mmbtc09,wvc08}. They
suggest that SPH may underestimate turbulence in the central region. 

\citet{ams07} carried out a series of experiments in order to compare
and contrast SPH and AMR in more of a ``controlled'' environment. They
conclude that there is a ``fundamental'' discrepancy between these
approaches, by demonstrating that SPH, at least in its conventional
form, cannot capture Kelvin-Helmholtz instabilities (KHI) as accurately
as an AMR approach. They further suggest that this discrepancy is not
due to one of resolution, but is a fundamental attribute of the scheme
itself \citep[see also][]{ii02,ojeqf03}.

We also note that \citet{vs10a} developed a moving mesh code, {\tt AREPO}, which combines the advantages of Lagrangian method and the superior hydrodynamics modelling of mesh codes.
Many comparison studies between SPH and the moving mesh code are seen in
\citet{kvssh12}, \citet{svksh12} and \citet{vsksh12}.

Recently, \citet{dp04,dp08,dp12} described a new scheme to improve the
conventional implementation of SPH and demonstrated the successful
capture of KHI \citep[see also][for alternative solutions]{rha10,ta11,rh12,mbbc11,gge12,sm12}. 
In what follows, we apply this scheme to our original
galactic chemodynamics code, {\tt GCD+} \citep{kg03a}.  {\tt GCD+} is a
three-dimensional tree $N$-body/SPH code that incorporates self-gravity,
hydrodynamics, radiative cooling, star formation, supernova feedback,
and metal enrichment.  
At its heart, the new scheme differentiates
itself from the conventional approach via the manner by which diffusion
of thermal energy is introduced; we adopt primarily the formalism
described by \citet{rp07}.
We demonstrate that this new scheme does
indeed advance the abilities of SPH codes in a suite of controlled hydrodynamics tests.
We focus only upon hydrodynamics simulations in this paper, and will study cases
including radiative cooling and star formation in a forthcoming work.
Note that our applied schemes have 
previously been presented in the literature, and as such, none of them 
are 'new'. However, {\tt GCD+} is a unique code, and we combined the advanced 
SPH schemes suggested by different researchers - i.e., the \citet{rp07} scheme,
the entropy equation by \citet{sh02}, and 
the \citet{sm09} time step limiter, which in consort, 
contribute to make {\tt GCD+} a new and advanced galaxy simulation code. There 
are certainly other extant SPH schemes which are more advanced, but 
within the context of galaxy simulations, this new version 
of {\tt GCD+} is somewhat unique. The present version of the code has been 
successfully applied to galaxy simulations \citep{rk12,gkc12b}
- i.e. we have confirmed that the same scheme described in 
this paper is applicable to galaxy simulations. This paper describes the 
performance of the updated {\tt GCD+} when applied to basic hydrodynamics 
tests. Note that our aim is not to test the code against extensive sets 
of such tests. We focus on only several tests useful for simulations of 
galaxy formation and evolution. We present how the code behaves in 
various situations, and how we chose the optimised parameter set of the 
new SPH scheme for our applications to galaxy simulations. We stress 
that this paper describes the performance for a new and practical galaxy 
simulation code, not for a specialised code which perhaps performs 
better than our code for some specific test simulations.

 Section~\ref{upver-sec} describes briefly the implementation of this new
scheme within {\tt GCD+}.
Section~\ref{res-sec} presents the performance of the new version of {\tt GCD+}  under several basic hydrodynamics tests. We here focus on the level of smoothing, i.e. number of neighbour particles, $N_{\rm nb}$, and several parameters involved in the artificial viscosity scheme.
A summary of this study is presented in Section~\ref{sum-sec}.

\section{{\tt GCD+} Update: Advancing Galactic Chemodynamics}
\label{upver-sec}

 We now describe the specific modifications made to the galactic
chemodynamics code {\tt GCD+}, which themselves are patterned closely
after the methodology described by \citet{rp07}.  As such, we only
outline the final formulae adopted, and refer the interested reader to
\citet{rp07} for their formal derivation.

The density of the $i$-th SPH particle is defined by 
%%%%%%%%%%   equation
\begin{equation}
 \rho_{i} = \sum_{j} m_j W(r_{ij},h_{i}).
\label{den-eq}
\end{equation}
\noindent
where $r_{ij}\equiv |\bmath{x}_i-\bmath{x}_j|$, and $h_i$ is the smoothing length of the $i$-th particle.
The SPH smoothing kernel of $W$ is described by a spherically symmetric spline kernel \citep{ml85,ms96},
%%%%%%%%%%   equation 1
\begin{equation}
\begin{array}{l}
W(r, h)=\frac{8}{\pi h^{3}} \nonumber \\
 \times \left\{ \begin{array}{cc}
 1-6(r/h)^{2}+6(r/h)^{3} & {\rm if}\ 0\leq r/h\leq 1/2, \nonumber \\
 2[1-(r/h)]^{3}      & {\rm if}\ 1/2\leq r/h\leq 1, \nonumber \\
 0               & {\rm otherwise}.
\end{array} \right.\\
\end{array} 
\label{w-eq}
\end{equation}
\noindent
We note in passing that the new version of {\tt GCD+} only takes into account the smoothing length
of the $i$-th particle, $h_i$, to derive the density, while the original version of {\tt GCD+} \citep{dk99} used 
the pair-averaged smoothing length, $h_{ij}=(h_i+h_j)/2$.
The smoothing length is determined by
%%%%%%%%%%   equation
\begin{equation}
 h_{i} = \eta \left(\frac{m_i}{\rho_i}\right)^{1/3}.
\label{h-eq}
\end{equation}
\noindent
Here $\eta$ is a free parameter; we compare the cases of $\eta=2$ and 2.4 in the next section. 
The solution of equation (\ref{h-eq}) is calculated iteratively until the relative change between two iterations
is smaller than $10^{-3}$ \citep[see][for more details]{pm07}.
Note that in our definition of the kernel,
our smoothing length corresponds to twice that used by \citet{rp07}, who adopt $\eta=1.2$.
We take this simple traditional kernel of equation (\ref{w-eq}), and do not consider more sophisticated kernels suggested recently by several authors \citep[e.g.][]{rha10,vdrrd10,rh12,da12}. Although there are many benefits of applying more sophisticated kernels, it is also demonstrated that such kernel are unstable when the number of neighbour particles is too low \citep[e.g.][]{da12}. We notice from our applications to galaxy evolution simulations \citep[e.g.][]{rk12}, that equation (\ref{h-eq}) leads to a lower number of neighbour particles around the density peak, compared to a nearly homogeneous density region. 
Therefore, in this paper we use the traditional kernel which is known to be more stable with a small number of neighbour particles.

Euler's equation is written as
%%%%%%%%%%   equation
\begin{eqnarray}
 \frac{d \bmath{v}_i}{dt} & = &
 -\sum_{j} m_j \left\{ \frac{P_i}{\Omega_i \rho_i^2} \nabla_i W_{ij}(h_i)
  +\frac{P_j}{\Omega_j \rho_j^2} \nabla_i W_{ij}(h_j) \right\} \nonumber \\
 & + & Q_{v,i} \nonumber \\
 & - & G\sum_j m_j\left\{
 \frac{\phi^{'}_{ij} (h_i)+\phi^{'}_{ij} (h_j)}{2} \right\} 
 \bmath{e}_{ij}
  \nonumber \\
 & - & \frac{G}{2} \sum_j m_j \left\{ \frac{\zeta_i}{\Omega_i} 
   \nabla_i W_{ij}(h_i) +\frac{\zeta_j}{\Omega_j} \nabla_i W_{ij}(h_j)\right\}.
\label{Eu-eq} 
\end{eqnarray}
\noindent
The first term of equation~(\ref{Eu-eq}) corresponds to the pressure gradient,
where $W_{ij}(h_i)=W(r_{ij},h_i)$, $\nabla_i W_{ij}(h_i)=\partial W(r_{ij},h_i)/\partial \bmath{x}_i$ and 
%%%%%%%%%%   equation
\begin{equation}
 \Omega_i=1-\frac{\partial h_i}{\partial \rho_i} \sum_k 
 \frac{\partial W_{ik}(h_i)}{\partial h_i}.
\label{Omg-eq} 
\end{equation}
\noindent
From equation (\ref{h-eq}), $\partial h_i/\partial \rho_i=-h_i/(3\rho_i)$. 
To mitigate the pairing instability \citep{ss81}, 
following \citet{tc92} and \citet{ms96}, we also apply the constant 
kernel gradient at $(r/h) \leq 1/3$, i.e.
%%%%%%%%%%   equation
\begin{equation}
 \nabla W = \nabla W(r/h=1/3)\ \ \ \ \ \ {\rm if}\ r/h\leq 1/3.
\label{dWc-eq}
\end{equation}
With a small value of $\eta$ (see eq.~\ref{h-eq}) applied in this paper, it is known that the 
pairing instability is not serious if the particles are homogeneously 
distributed \citep[e.g.][]{dp12}. 
However, in galaxy simulations we often have a higher number of neighbour particles than that expected in the 
homogeneous case. {\bf Also, because of thermal instability due to radiative cooling, the minimum smoothing and 
softening are often required to be applied \citep[e.g.][]{nw93,wsq04}\footnote{
For example, we often set the minimum smoothing length to be half that of the minimum softening length.
When the smoothing length reaches the minimum value, we set $\Omega_i=1$ (eq.~\ref{Omg-eq}). We also
set $\zeta_i=0$ (eq. \ref{zeta-eq}), when the softening length hits the minimum value.
In this paper, we do not apply the minimum softening or smoothing lengths.}, which could enhance the pairing instability.}
Therefore we apply equation (\ref{dWc-eq}) to 
only the pressure gradient for safety. This 
could be a problematic choice because in theory this breaks the 
consistency between the kernel and the kernel derivative. However, the 
results in Section \ref{res-sec} encouragingly demonstrate that our applied scheme 
works well for the hydrodynamics tests.

The second term of equation~(\ref{Eu-eq}) corresponds to the artificial 
viscosity (AV), 
%%%%%%%%%%   equation
\begin{eqnarray}
 Q_{v,i} & =& - \left( \sum_{j} m_j \frac{\alpha_{ij}^{\rm AV} (t) v_{\rm sig} 
    \bmath{v}_{ij} \cdot \bmath{e}_{ij}}{\rho_{ij}} \right)\nonumber \\
 &\times & \overline{\nabla_i W_{ij}}
  \ \ \ \ \ \ \ {\rm if\ } \bmath{x}_{ij}\cdot\bmath{v}_{ij}<0 \nonumber \\
 & = & 0\ \ \ \ \ \ \ \ \ \ \ \ \ \ \rm otherwise,
\label{qv-eq}
\end{eqnarray}
\noindent
where $\bmath{v}_{ij}=\bmath{v}_i-\bmath{v}_j$,
$\bmath{e}_{ij}=(\bmath{x}_i-\bmath{x}_j)/|\bmath{x}_i-\bmath{x}_j|$,
$\rho_{ij}=(\rho_i+\rho_j)/2$ and
\begin{equation}
\overline{\nabla_i W_{ij}} = 
 \frac{1}{2} \left\{\frac{1}{\Omega_i} \nabla_i W_{ij}(h_i)
   + \frac{1}{\Omega_j} \nabla_i W_{ij} (h_j)\right\}.
\end{equation}
\noindent
The signal velocity $v_{\rm sig}$ adopted is
%%%%%%%%%%   equation
\begin{equation}
  v_{\rm sig}=\frac{c_{s,i}+c_{s,j}- \beta^{\rm AV}\bmath{v}_{ij}\cdot\bmath{e}_{ij}}{2},
\label{vsig-eq} 
\end{equation}
\noindent
where $c_{s,i}$ is the sound velocity of the $i$-th particle. We set $\beta^{\rm AV}=3.0$ as explained later. 
The amount of AV is controlled by 
a time-dependent parameter,
%%%%%%%%%%   equation
\begin{equation}
 \alpha^{\rm AV}_{ij}(t)=\frac{1}{4}
  (\alpha^{\rm AV}_i(t)+\alpha^{\rm AV}_j(t))
  (f_i+f_j),
\label{alav-eq} 
\end{equation}
\noindent
where \citep{db95}
%%%%%%%%%%   equation
\begin{equation}
  f_i = 
  \frac{|\langle {\bf \nabla}\cdot \bmath{v}\rangle_i|}
       {|\langle {\bf \nabla}\cdot \bmath{v}\rangle_i|
       +|\langle {\bf \nabla}\times\bmath{v}\rangle_i|
       +0.0002 c_{{\rm s},i}/h_i},
\label{fi-eq}
\end{equation}
\begin{equation}
 \langle\nabla \cdot \mbox{\boldmath $v$} \rangle_i
   =  -\frac{1}{\rho_i}
 \sum_{j} m_j \mbox{\boldmath $v$}_{ij}
 \cdot \nabla_i W_{ij}(h_{i}), 
\end{equation}
and
\begin{equation}
\langle \nabla \times \mbox{\boldmath $v$} \rangle_{i, x}
 =  -\frac{1}{\rho_i} \sum_{j} m_j
  \left[v_{ij,z}
 \nabla_{i,y} W_{ij}(h_{i})
- v_{ij,y}
 \nabla_{i,z} W_{ij}(h_{i}) \right],
\end{equation}
\noindent
in order to suppress AV in pure shear flows.
The viscous parameter $\alpha^{\rm AV}_i(t)$ varies with time.
\citet{jm97} suggested 
the following function to evolve this viscous parameter
\citep[see also][for a more sophisticated AV switch]{cd10}:
%%%%%%%%%%   equation
\begin{equation}
  \frac{d \alpha^{\rm AV}_{i}(t)}{dt} = 
-\frac{\alpha^{\rm AV}_i(t)-\alpha^{\rm AV}_{\rm min}}{\tau_i}+S_i,
\label{dadt-eq}
\end{equation}
\noindent
where we set $\alpha^{\rm AV}_{\rm min}=0.5$ or 0.05, depending on $\eta$, which will be discussed below, and
%%%%%%%%%%   equation
\begin{equation}
  \tau_i=\frac{h_i}{0.2 c_{s,i}}.
\label{tauav-eq}
\end{equation}
\noindent
\citet{rdtp00} and \citet{rp07} adopt the source term,
%%%%%%%%%%   equation
\begin{equation}
  S_i=\max(-\nabla_i\cdot\bmath{v}_i,0) (\alpha^{\rm AV}_{\rm max}-\alpha^{\rm AV}_i(t)),
\label{savrp07-eq}
\end{equation}
\noindent
and set the maximum of the viscous parameter to be $\alpha^{\rm AV}_{\rm max}=2.0$.

The third term of equation~(\ref{Eu-eq}) corresponds to the gravitational force, and employs the adaptive gravitational force softening suggested in \citet{pm07}, where the softening length is matched to that of the smoothing length.
The fourth term of equation~(\ref{Eu-eq}) is
the correction term for adaptive softening, where 
%%%%%%%%%%   equation
\begin{equation}
 \zeta_i=\frac{\partial h_i}{\partial \rho_i} \sum_j m_j 
  \frac{\partial \phi_{ij}(h_i)}{\partial h_i}.
\label{zeta-eq} 
\end{equation}
\noindent
We apply a cubic splice softening, as suggested by \citet{pm07};
the associated formulae for
$\phi^{'}$ and $\partial \phi/\partial h$ can also be 
found in their paper.

 Following \citet{sh02} \citep[and different from][]{rp07}, instead of the energy equation, we follow the entropy equation, which is written as
%%%%%%%%%%   equation
\begin{eqnarray}
 \frac{d A_i}{dt} & = & \frac{\gamma-1}{\rho^{\gamma-1}}Q_{u,ij},
\label{ene-eq} 
\end{eqnarray}
\noindent
where $A_i=P_i/\rho_i^{\gamma}=((\gamma-1)/\rho_i^{\gamma-1})u_i$ is entropy and $u_i$ is the thermal energy hereafter, of the $i$-th particle.
$Q_{u,ij}$ is zero if $\bmath{x}_{ij}\cdot\bmath{v}_{ij}>0$.
Otherwise, it is described by
%%%%%%%%%%   equation
\begin{eqnarray}
 Q_{u,ij} & = & -\sum_j \frac{m_j v_{\rm sig}}{\rho_{ij}}
  \left\{ \frac{\alpha^{\rm AV}_{ij}(t)}{2} 
  (\bmath{v}_{ij}\cdot\bmath{e}_{ij})^2
  -\alpha^{\rm C}_{ij}(t) (u_i-u_j)\right\}  \nonumber \\
  & \times & \bmath{e}_{ij}\cdot\overline{\nabla_i W_{ij}},
\label{qu-eq} 
\end{eqnarray}
\noindent
where $\alpha^{\rm C}_{ij}(t)=(\alpha^{\rm C}_i(t)+\alpha^{\rm C}_j(t))/2$.
The second term within the parentheses of equation~(\ref{qu-eq}) corresponds
to the artificial thermal conductivity (AC) \citep{rp07,dp08}.
The thermal conductivity parameter, $\alpha^{\rm C}$, evolves
between $\alpha^{\rm C}=0$ and 2 following 
%%%%%%%%%%   equation
\begin{equation}
 \frac{d \alpha^{\rm C}_{i}(t)}{dt} = -\frac{\alpha^{\rm C}_i(t)}{\tau_i}
  +S^{\rm C}_i,
\label{dacdt-eq} 
\end{equation}
\noindent
where the source term is 
%%%%%%%%%%   equation
\begin{equation}
 S^{\rm C}_i=0.05 h_i |\nabla^2 u_i|/\sqrt{u_i},
\label{scc-eq} 
\end{equation}
\noindent
and \citep{lb85}
%%%%%%%%%%   equation
\begin{equation}
 \nabla^2 u_i=2 \sum_j m_j \frac{u_i-u_j}{\rho_j} 
  \frac{\bmath{e}_{ij}\cdot\overline{\nabla_i W_{ij}}}{r_{ij}}.
\label{d2u-eq} 
\end{equation}
%

%For collisionless N-body particles (such as DM),
%only the gravitational force term of equation~(\ref{Eu-eq}) 
%is taken into account, as follows:
%%%%%%%%%%%   equation
%\begin{eqnarray}
% \frac{d \bmath{v}_i}{dt} & = &
%  - G\sum_j m_j\left\{
% \frac{\phi^{'}_{ij} (h_i)+\phi^{'}_{ij} (h_j)}{2} \right\} 
% \bmath{e}_{ij}
%  \nonumber \\
% & - & \frac{G}{2} \sum_j m_j \left\{ \frac{\zeta_i}{\Omega_i} 
%   \nabla_i W_{ij}(h_i) +\frac{\zeta_j}{\Omega_j} \nabla_i W_{ij}(h_j)\right\}.
%\label{Eunbody-eq} 
%\end{eqnarray}
%% 
%\noindent
%The adaptive softening length of the DM is calculated 
%using equation~(\ref{h-eq}) and for the DM we apply $\eta_{\rm DM}=4$,
%as the simulations under consideration generally have DM particles of
%higher mass.
%Note that for the DM we use the mass density, 
%$\rho_{i,{\rm DM}}=\Sigma_j m_{j,{\rm DM}} W(r_{ij},h_i)$, in 
%equation~(\ref{h-eq}). Therefore, this higher $\eta_{\rm DM}$ 
%results in a DM softening length which is larger than that for
%the gas when the number density of DM particles is comparable to that of
%the gas.
%We also set the minimum softening length $\epsilon_{\rm min}$ 
%for both N-body and SPH particles to be
%$\epsilon_{\rm min}=(c_{\epsilon} m_p(M_{\sun}))^{1/3}$~pc,
%where $m_p$ is the mass of the particles.
%For the work described here, 
%we employ $c_{\epsilon}=3000$, which results in 
%similar softening to that we used in our previous work
%\citep{kg03a,kg03b}.
%When $h_i=\epsilon_{min}$, we set $\Omega_i=1$ and $\zeta_i=0$.

We apply an individual timestep scheme to
integrate equations~(\ref{Eu-eq}) and (\ref{ene-eq}).
We also employ the timestep limiter suggested by \citet{sm09}.
In Section \ref{ple-sec} we demonstrate that this timestep limiter is critical. 
The timestep for SPH particles is based upon 
$dt_i=\min(dt_{{\rm CFL},ij},dt_{{\rm DYN},i})$, where the
Courant-Friedrich-Levy condition is calculated by 
%%%%%%%%%%   equation
\begin{equation}
  dt_{{\rm CFL},ij}=C_{{\rm CFL}} \frac{0.5 h_i}{v_{dt,ij}},
\label{dtcfl-eq} 
\end{equation}
\noindent
where $v_{dt,ij}=v_{{\rm sig},ij}$ if $\bmath{x}_{ij}\cdot\bmath{v}_{ij}<0$, 
otherwise $v_{dt,ij}=
0.5 (c_{s,i}+c_{s,j}-\bmath{v}_{ij}\cdot\bmath{e}_{ij})$.
We set $C_{\rm CFL}=0.2$.
The requirement that the force should not change significantly
within one timestep is satisfied by 
%%%%%%%%%%   equation
\begin{equation}
  dt_{{\rm DYN},i}=C_{{\rm DYN}} 
   \left(\frac{0.5 h_i}{|d \bmath{v_i}/dt|}\right)^{1/2}.
\label{dtdyn-eq} 
\end{equation}
\noindent
We set $C_{\rm DYN}=0.2$. The values of $C_{\rm CFL}$ and $C_{\rm DYN}$ are chosen
after testing in one-dimensional Riemann problems in Section \ref{stub-sec}.
We integrate equation (\ref{Eu-eq}) with the leap-frog method, and equation (\ref{ene-eq}) 
using the trapezoidal rule \citep{hk89}.
We also implement the FAST scheme \citep{sm10} which allows the use of different timesteps for integrating hydrodynamics and gravity.

%%% stub A 

\begin{figure}
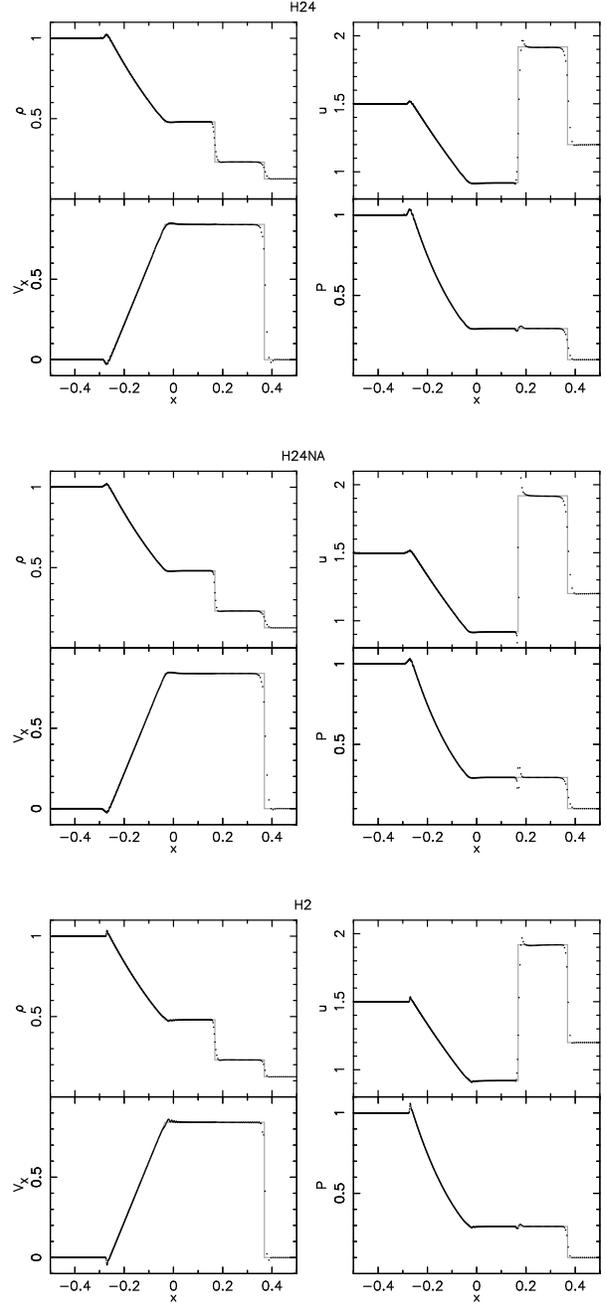

\centering
\includegraphics[width=\hsize]{f1a.ps}\\
\includegraphics[width=\hsize]{f1b.ps}\\
\includegraphics[width=\hsize]{f1c.ps}
\caption{
Results of a one-dimensional Riemann problem A with models H24 (top), H24NA (middle) and H2 (bottom) at $t=0.2$. 
The grey line represents the analytic solution.
}
\label{stA-fig}
\end{figure}

\begin{figure}
\centering
\includegraphics[width=\hsize]{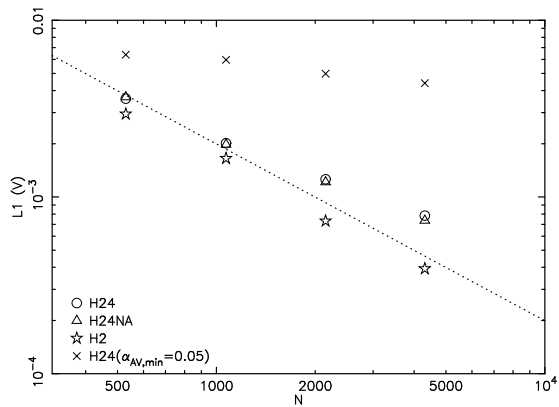}
\caption{
 Comparison of $L1(V)$ of from different models with different number of particles {\bf a one-dimensional Riemann problem A}. Circle, triangle and stars show the results of models H24, H24NA and H2. Cross shows model H24 but applying $\alpha_{\rm AV,min}=0.05$. The dashed line indicate $L1(V)\propto N^{-1}$ relation for a reference. 
}
\label{stAl1v-fig}
\end{figure}

\section{Results}
\label{res-sec}

 Having outlined the improvements made to {\tt GCD+}, we now test its performance. We especially explore the impact of the choice of the parameter $\eta$ of equation~(\ref{h-eq}) and the AC and AV switch that are newly implemented. We present the results of mainly three different models:
models with $\eta=2.4$ (H24) and 2 (H2) and a model with $\eta=2.4$ without the AC or the AV switch (H24NA).  A summary of these models is presented in Table~\ref{mod-tab}. Models H24 and H2 are expected to have neighbouring number of particles of $N_{\rm nb}\sim58$ and $\sim33$ respectively, when the particles are distributed homogeneously in three dimensional space. These are conventionally used values. We did not take a higher value of $\eta$, because it leads to a larger number of neighbour particles, and requires more computational costs\footnote{Using a large number of neighbour particles are also not recommended with the traditional SPH kernel of equation (\ref{w-eq}), because of the increasing the pairing instability \citep{ss81,dp12,da12}. Although we use equation (\ref{dWc-eq}) to mitigate the paring instability, we avoid applying a large $\eta$ also for this reason.}.
We applied a higher $\alpha_{\rm AV,min}$ for model H24. The reason behind this choice is demonstrated in Section \ref{stub-sec}. We apply $\beta^{\rm AV}=3.0$ in all the models. We demostrate in Section \ref{ple-sec} that a lower $\beta^{\rm AV}$ fails to reproduces the analytic solution of the point-like explosion test  \citep[see also][who recommended an even higher value of $\beta^{\rm AV}=4.0$]{pf10}.

%%% Tab 1n

\begin{table}
 \centering
 \begin{minipage}{\hsize}
 \caption{Model Parameters}
 \label{mod-tab}
 \begin{tabular}{@{}ccccc}
\hline
Model & $\eta$ & $\alpha_{\rm AV,min}$ & AC & AV switch \\
 \hline
 H24 & 2.4 & 0.5 & yes & yes \\
 H24NA & 2.4 & 1.0 & no & no \\
 H2  & 2.0 & 0.05 & yes & yes \\
\hline
 \end{tabular}
 \end{minipage}
\end{table}

%%% Tab 2

\begin{table}
 \centering
 \begin{minipage}{\hsize}
 \caption{Riemann problems ($\gamma=5/3$) initial conditions.}
 \label{rp-tab}
 \begin{tabular}{@{}ccccccc}
\hline
Problem & $\rho_{\rm L}$ & $V_{\rm L}$ & $P_{\rm L}$ 
& $\rho_{\rm R}$ & $V_{\rm R}$ & $P_{\rm R}$ \\
 \hline
 A & 1.0 & 0.0 & 1.0 & 0.125 & 0.0 & 0.1 \\
 B & 1.0 & $-2.0$ & 0.4 & 1.0 & 2.0 & 0.4 \\
 C & 1.0 & 0.0 & 1000.0 & 1.0 & 0.0 & 0.01 \\
\hline
 \end{tabular}
 \end{minipage}
\end{table}

%%% stub B 

\begin{figure}
\centering
\includegraphics[width=\hsize]{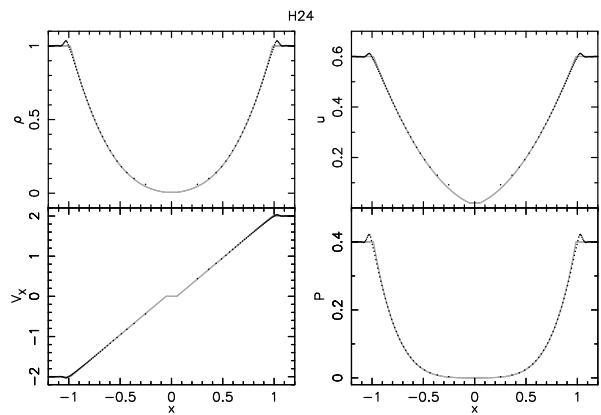}
\caption{
Results of a one-dimensional Riemann problem B with model H24 at $t=0.35$. 
The grey line represents the analytic solution.
}
\label{stBe24-fig}
\end{figure}

\begin{figure}
\centering
\includegraphics[width=\hsize]{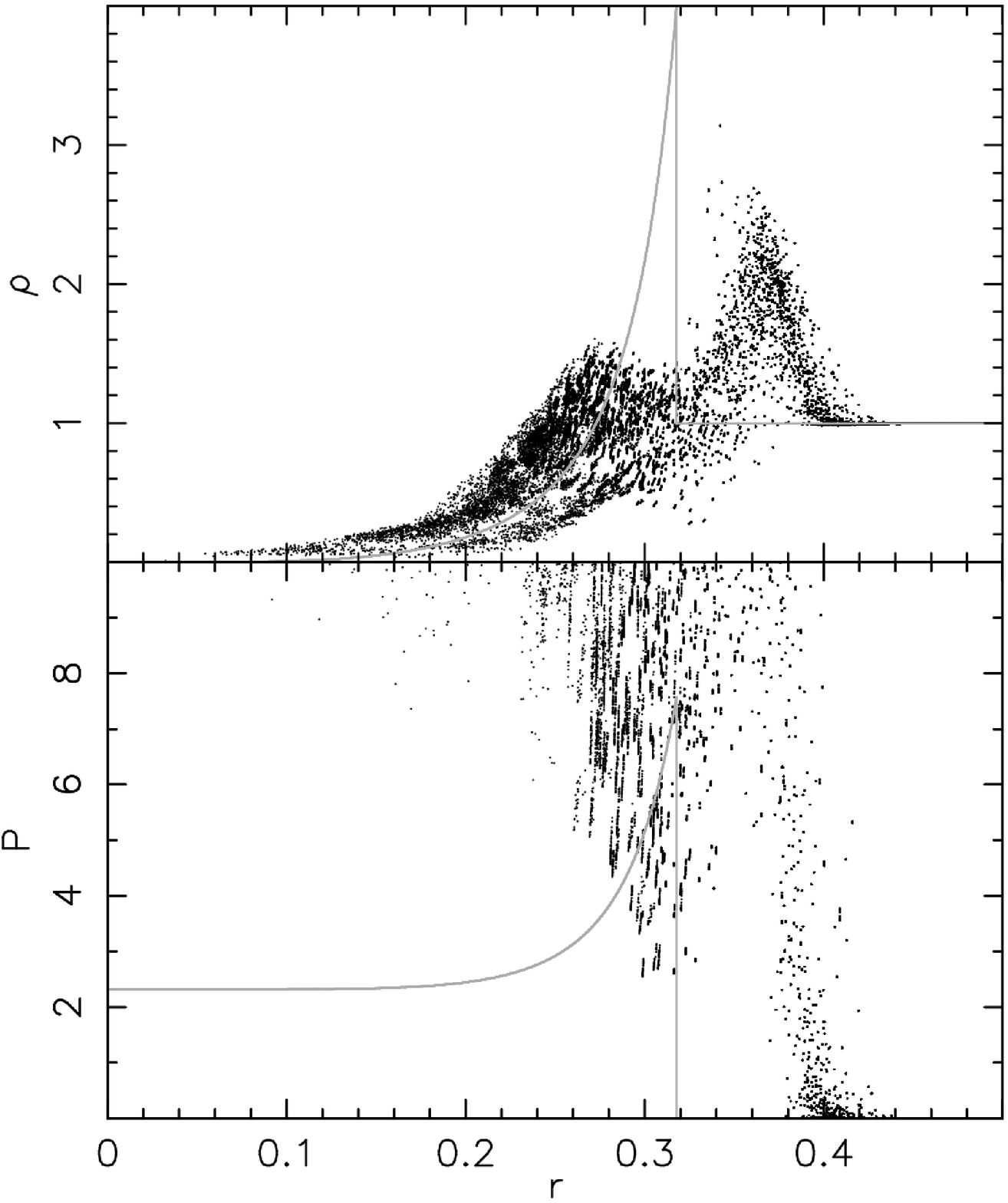}
\caption{
Comparison of $L1(V)$ from different models with different number of particles for {\bf a one-dimensional Riemann} problem B. The dashed line indicate $L1(V)\propto N^{-1}$ relation for a reference. Symbols as defined in Fig. \ref{stAl1v-fig}.
}
\label{stBl1v-fig}
\end{figure}

%%% stub C

\begin{figure}
\centering
\includegraphics[width=\hsize]{f5a.ps}\\
\includegraphics[width=\hsize]{f5b.ps}\\
\includegraphics[width=\hsize]{f5c.ps}
\caption{
Results of a one-dimensional Riemann problem C with models H24 (top), H24NA (middle) and H2 (bottom) at $t=0.008$. 
The grey line represents the analytic solution.
}
\label{stC-fig}
\end{figure}

\begin{figure}
\centering
\includegraphics[width=\hsize]{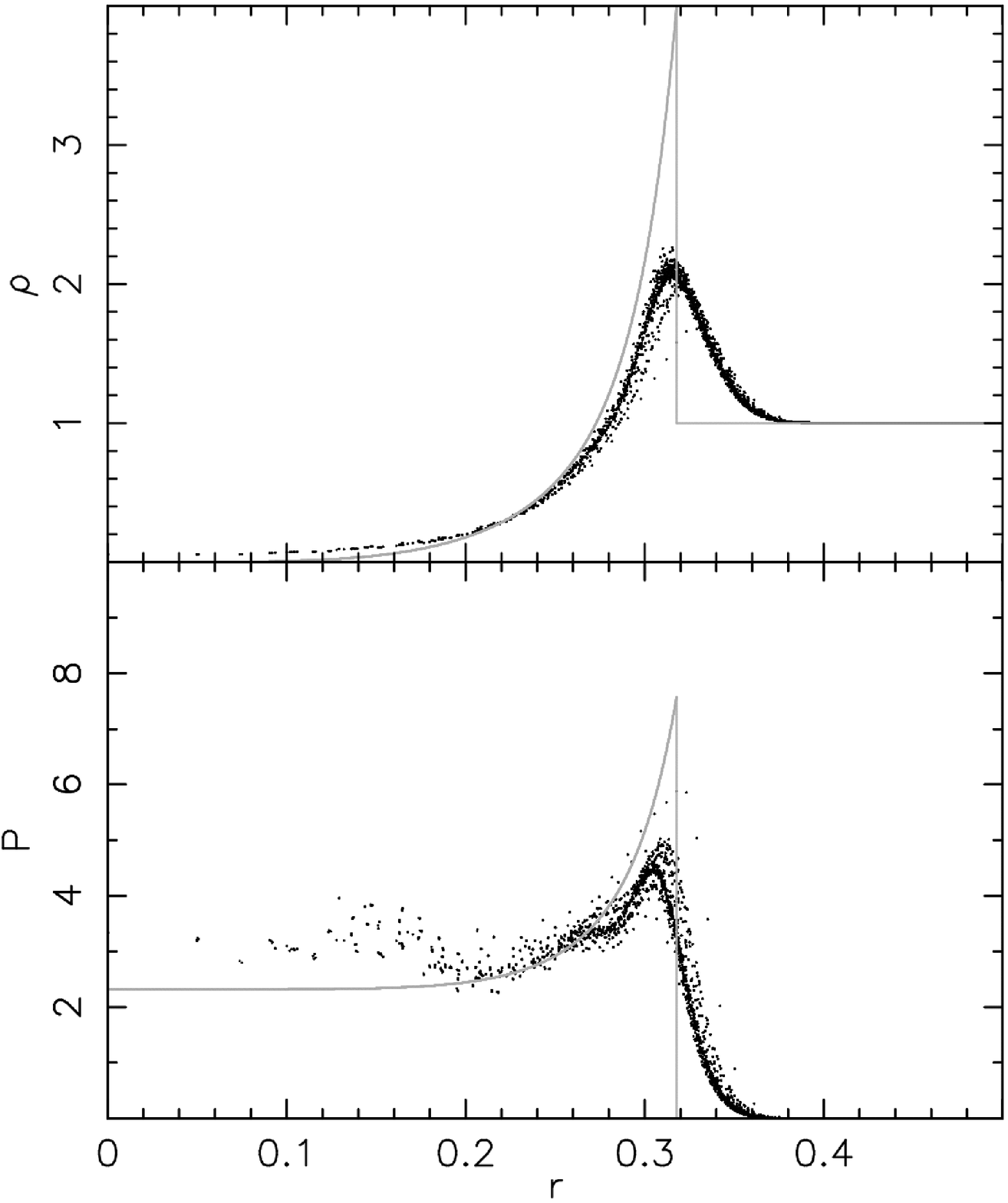}
\caption{
Comparison of $L1(V)$ from different models with different number of particles for {\bf a one-dimensional Riemann} problem C. The dashed line indicate $L1(V)\propto N^{-1}$ relation for a reference. Symbols as defined in Fig. \ref{stAl1v-fig}.
}
\label{stCl1v-fig}
\end{figure}

%%% 1 June

\subsection{One-dimensional Riemann Problems}
\label{stub-sec}

Our first experiments involve a version of the
classical one-dimensional Riemann problems \citep[e.g.][]{et97}.  The initial conditions
are set by assuming the simulation region spans from 
$x=-0.5$ to $x=0.5$; the region for which $x<0$ is set to $(\rho_{\rm L},P_{\rm L},V_{\rm L})$,
and the region for which $x>0$ is set to $(\rho_{\rm R},P_{\rm R},V_{\rm R})$, adopting $\gamma=5/3$ throughout. We show three problems summarised in Table~\ref{rp-tab}.
Fig.~\ref{stA-fig} shows the results of problem A for models H24, H24NA and H2, using 540 particles. As also demonstrated in the literatures \citep[e.g.][]{dp08}, one can see a clear jump in thermal energy and pressure at the contact discontinuity in model H24NA. On the other hand, including the AC, the contact discontinuity is resolved, and the pressure and thermal energy distribution is much smoother in models H24 and H2, although there is still a small jump. A smooth pressure distribution at the contact discontinuity is key to accurately simulating KHI \citep{dp08}; as such, it would appear that models with the AC and AV switch are promising tools for modeling KHI within an SPH framework.
It is also remarkable that the number of particles employed to resolve the shock front in model H2 is so low.
As expected, if we adopt a higher $\eta$ value, a greater number of particles are required to resolve the shock front.

Following \citet{vs10b}, we measure an L1 error norm defined by
\begin{equation}
 L1(V)=\frac{1}{N}\sum_i|V_i-V_{\rm c}(x_i)|,
\end{equation}
where $N$ is the number of SPH particles, $V_i$ is the velocity of the particle $i$ and $V_{\rm c}(x_i)$ is the analytic solution for the problem. We run problem A with different models and different resolutions and summarise $L1(V)$ values in Fig.~\ref{stAl1v-fig}. Model H2 shows the error declining as 
$L1(V)\propto N^{-1}$, similar to what is shown in \citet{vs10b}. Interestingly, adopting higher $\eta$ leads to higher error and slower convergence. Although snapshot of model H24NA shows significantly worse results than model H24 (Fig.~\ref{stA-fig}), $L1(V)$ shows similar results. It means that $L1(V)$ is not a good measure for how well the code captures the contact discontinuity. We also show the $L1(V)$ results of a model with $\eta=2.4$ and $\alpha_{\rm AV, min}=0.05$ with crosses in Fig.~\ref{stAl1v-fig}. This clearly demonstrates that if we apply the lower $\alpha_{\rm AV, min}$ with a larger $\eta$, it induces unacceptably high scatter in the velocity field. Therefore, we apply the higher value of $\alpha_{\rm AV, min}$ for model H24.

Fig.~\ref{stBe24-fig} shows the results of problem B for model H24, using 540 particles. We do not show the results of each model for this problem, because all the models reproduce the analytic solution equally well. Fig.~\ref{stBl1v-fig} shows the $L1(V)$ results of problem B. As expected, since model H2 has less smoothing, the $L1(V)$ error norm is lower than models H24 and H24NA.

 Fig.~\ref{stC-fig} shows the results of problem C at $t=0.008$ for models H24, H24NA and H2. Fig.~\ref{stCl1v-fig} presents $L1(V)$ for problem C. There is less difference among the three different models in $L1(V)$ for problem C.  However, one can see a much bigger jump in thermal energy and pressure at the contact discontinuity in model H24NA, compared to problem A (Fig.~\ref{stA-fig}). Even for this stronger shock case, the AC helps to resolve the contact discontinuity and capture the correct shock feature. 
 
%%% Gresho

\begin{figure*}
\centering
\includegraphics[width=\hsize]{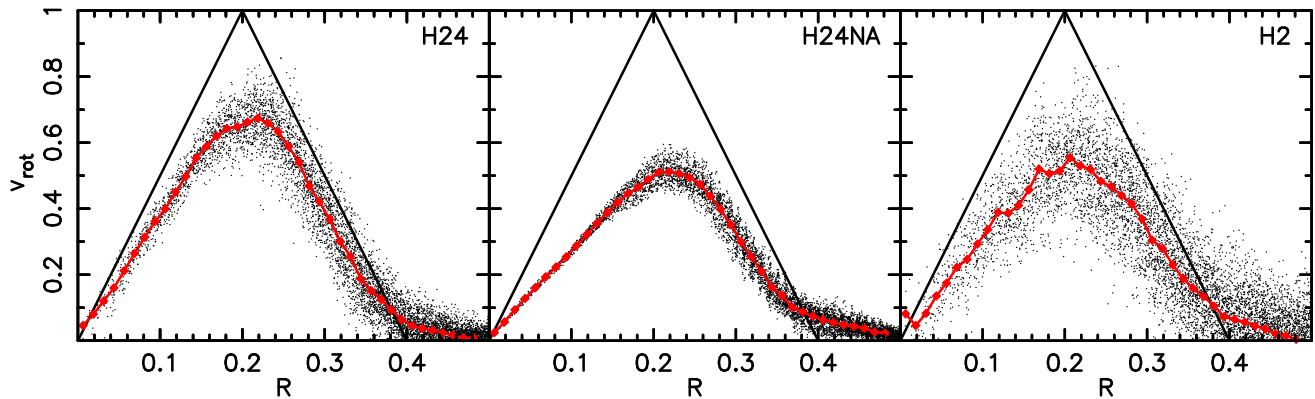}
\caption{
 Velocity profile at $t=1.0$ in the Gresho vortex test with 9040 particles which 
were initially set on a hexagonal grid (80 particles along x-axis). From left to right, the panels show the results of models H24, H24NA and H2. Red dots and lines show the mean values. The solid lines are the correct solution.
}
\label{greshon80-fig}
\end{figure*}

\subsection{Gresho Vortex Test}
\label{gresho-sec}

 To check the stability of our models in a rotating system, we run the so-called Gresho vortex test \citep{gc90,vs10b} with different models and different numbers of particles. This is a two-dimensional problem. We initially set particles on a hexagonal grid \citep{dp04} in a two-dimensional periodic region, and the rotation velocity as a function of radius as follows
%%%%%%%%%%   equation
\begin{equation}
V_{\rm rot}(R) = \left\{ \begin{array}{cc}
 5 R & {\rm for}\ 0\leq R\leq 0.2, \\
 2-5R     & {\rm for}\ 0.2\leq R \leq 0.4, \\
 0               & {\rm for}\ R \geq 0.4.
\end{array} \right. 
\label{greshovrot-eq}
\end{equation}
The gas density is constant, $\rho=1$, and $\gamma=5/3$ is adopted. We then assume an initial pressure, which is a function of radius, following
%%%%%%%%%%   equation
\begin{equation}
P(R) = \left\{ \begin{array}{cc}
 5+\frac{25}{2} R^2 & {\rm for}\ 0\leq R\leq 0.2, \\
 9+\frac{25}{2} R^2 & \\
 -20 R+4 \ln (R/0.2)    & {\rm for}\ 0.2\leq R \leq 0.4, \\
 3+4 \ln 2               & {\rm for}\ R \geq 0.4.
\end{array} \right. 
\label{greshop-eq}
\end{equation}
In this condition, the centrifugal force is balanced by the pressure gradient, and the initial rotation velocity should be maintained. 
  
\begin{figure}
\centering
\includegraphics[width=\hsize]{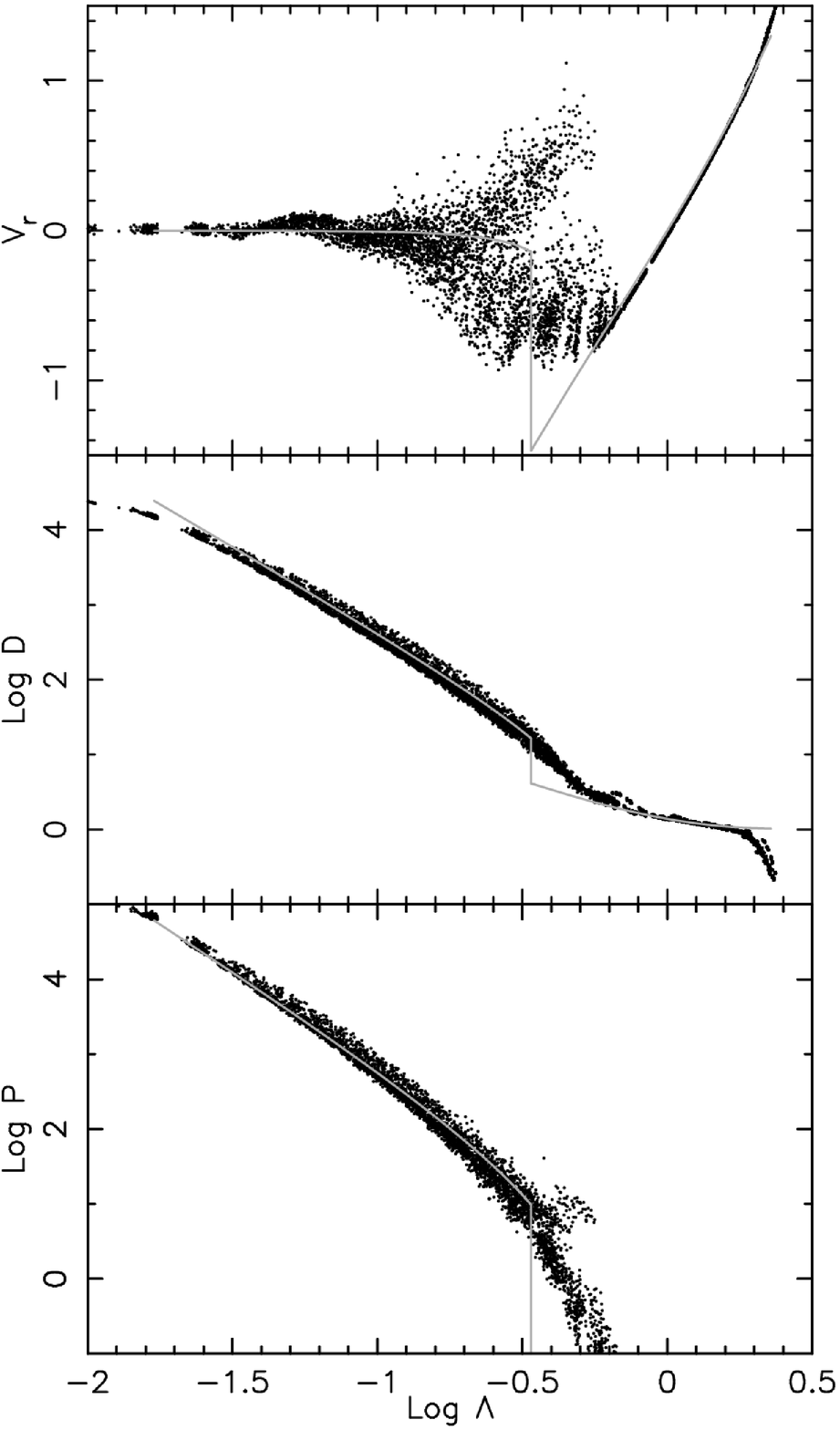}
\caption{
Comparison of $L1(V)$ from different models with different number of particles for Gresho vortex test. N indicates number of particle along radius within $R=0.5$. The dashed line indicate $L1(V)\propto N^{-1}$ relation for a reference. Symbols as defined in Fig. \ref{stAl1v-fig}.
}
\label{greshol1v-fig}
\end{figure}

 Fig.~\ref{greshon80-fig} shows the rotation velocity profile of all three models at $t=1.0$ in our lowest resolution test. Although the rotation velocity should be kept constant, all the models have slower rotation velocities at $t=1.0$ because of the angular momentum transfer due to the AV. Fig.~\ref{greshol1v-fig} displays $L1(V)$ error norm for the Gresho test with different resolutions. As also shown in \citet{vs10b}, all of our models show very slow convergence or saturation, i.e. higher resolution simulations do not improve the results significantly. Nevertheless, $L1(V)$ of model H24 is significantly lower than the other models. This demonstrates that the angular momentum transfer due to the AV is suppressed by applying larger $\eta$ and employing the AC and the AV switch. \citet{da12} discuss that adopting a more sophisticated kernel function will reduce $L1(V)$ error norm dramatically. However, at this stage we hesitate to use such kernels because of the possible instability when the number of the neighbour particles becomes low, as discussed above.

%%% Sedov r vs rho plot

\begin{figure*}
\centering
\includegraphics[width=0.45\hsize]{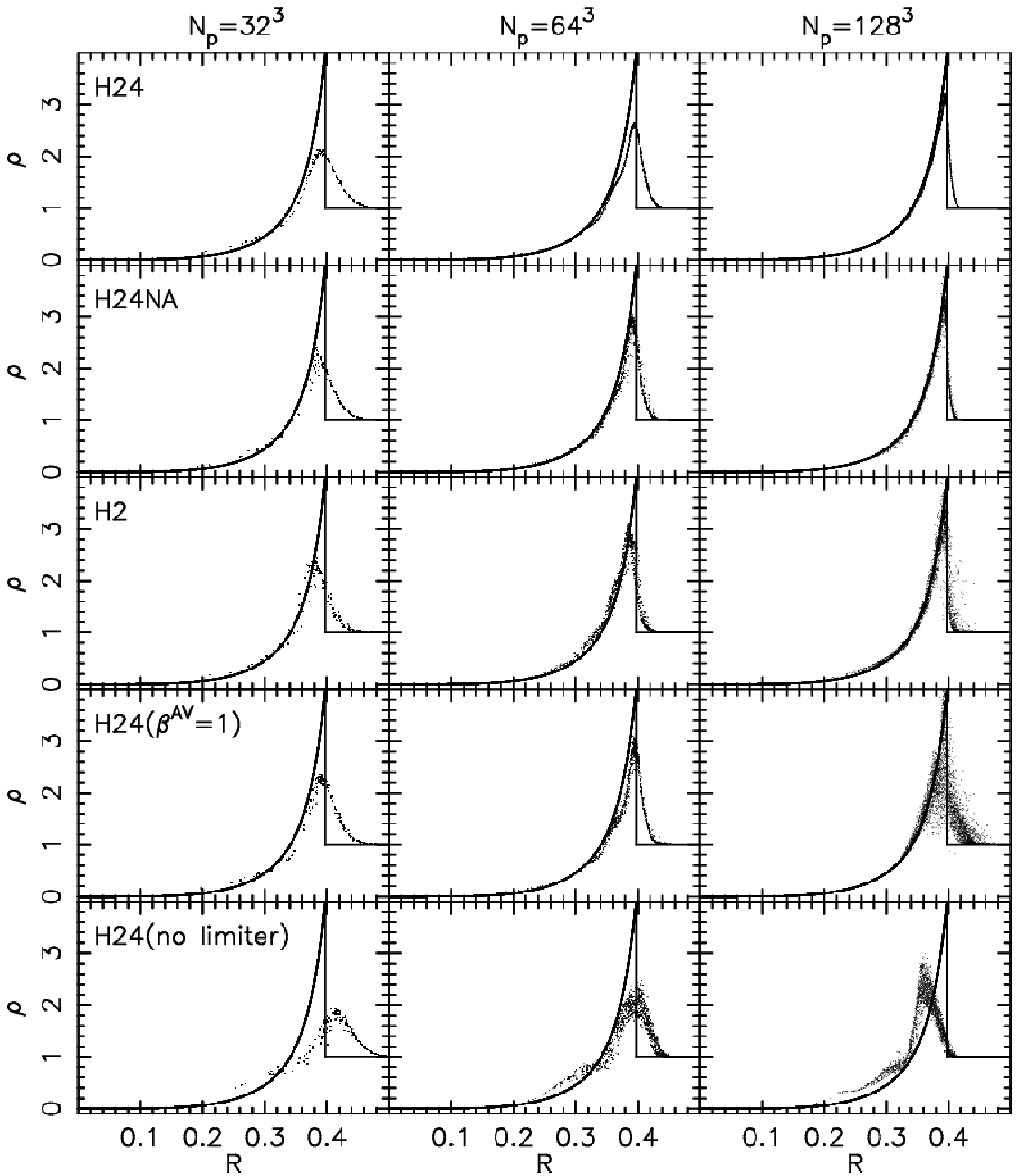}
\includegraphics[width=0.45\hsize]{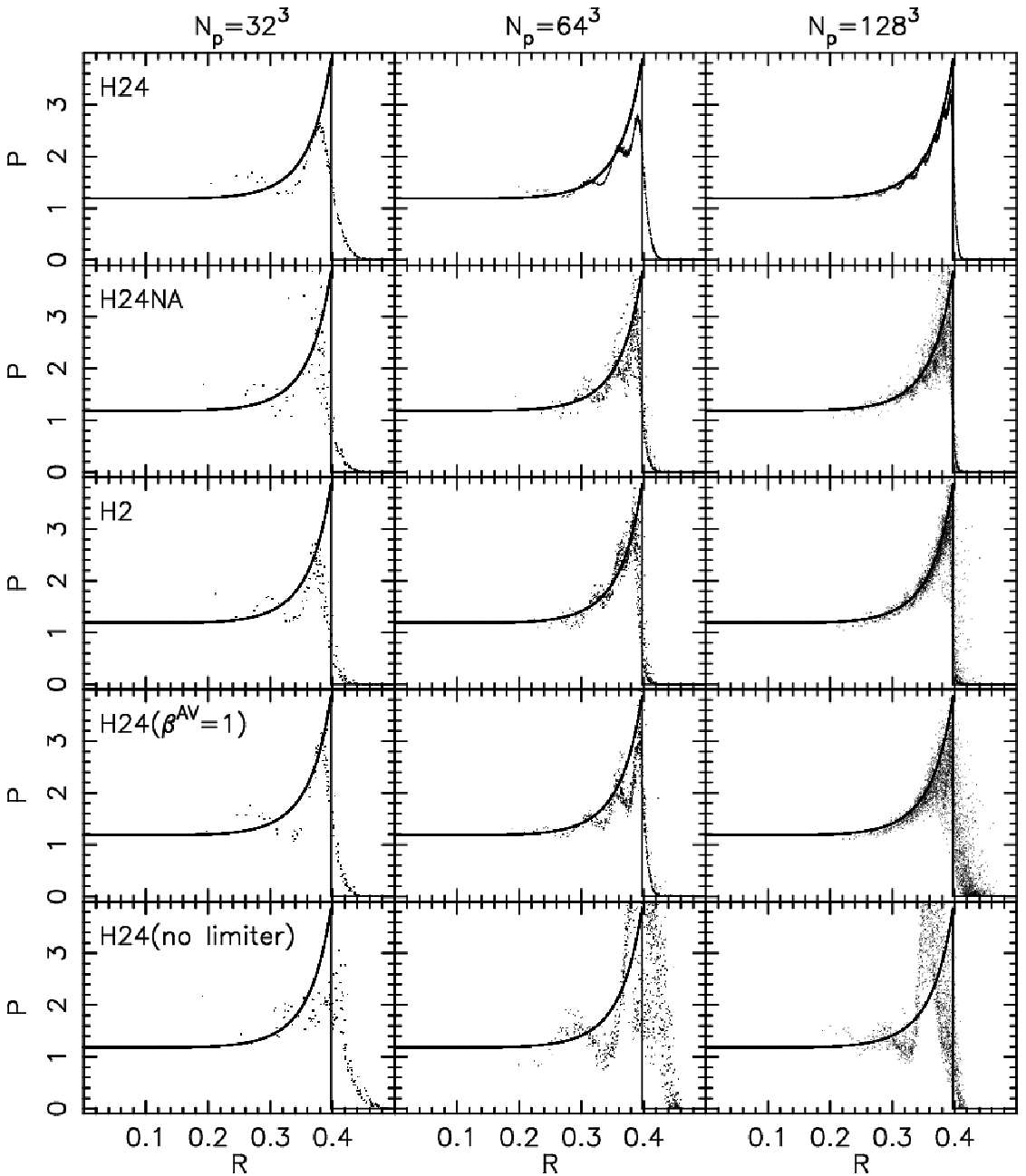}
\caption{
Radial density (left) and pressure (right) distributions at $t=0.07$ in the point-like explosion test with different models and resolutions. 
Left, middle and right panels show the results of $N_{\rm p}=32^3$, $64^3$ and $128^3$, and from top to bottom the panels present the results of models H24, H24NA, and H2, a model same as H24, but with $\beta^{\rm AV}=1$, and a model same as H24, but without the individual timestep limiter \citep{sm09}. The solid lines show the analytic solution. Note that we do not plot the particles in the region of $|x|<0.1, |y|<0.1$ and $|z|<0.1$, because the particles in these regions show incorrect behavior, possibly due to the initial grid particle setting. We also plot only every about $(N/32^3)$ particles.
}
\label{sedovr-fig}
\end{figure*}

\subsection{Point-like explosion test}
\label{ple-sec}

 We next consider the Sedov-Taylor-type spherical explosion test. Following \citet{sh02}, 
we set a three-dimensional periodic boundary box with a low-temperature and
homogeneous density ($\rho=1$). At $t=0$, we deposit $E=1$ energy on the central particle and simulate the evolution thereafter. The analytic solution can then be derived via the adoption of Sedov-Taylor self-similarity. Fig.~\ref{sedovr-fig} shows the density and pressure of the gas particles as a function of radius at $t=0.07$ for different models and different resolutions, while the solid line represents the analytic solution. We notice that the particles in the region of $|x|<0.1, |y|<0.1$ and $|z|<0.1$ show incorrect behavior, and do not plot them in Fig.~\ref{sedovr-fig}. We think that this is due to our initial setting of the particles at square grid points. The particles along each axis are in the special location, and the particles are aligned to the radial direction. However, it will be extremely rare in a galaxy simulation that many particles are radially aligned from a single star particle which is producing some feedback, like supernovae. For the purpose of calibrating the parameter for the galaxy simulations, we ignore such special condition in this test. 

Fig.~\ref{sedovr-fig} demonstrates that model H24 reproduces the analytic function well, and higher resolution simulations recover its analytic solution better. Model H24NA is equally good in density distribution. However, the pressure distribution shows a significant scatter. This demonstrates the importance of checking the pressure profile, in addition to the radial density profile. Model H2 shows a sharper density profile than model H24, however both the density and pressure display greater scatter. 
 In these figures, we also demonstrate that $\beta^{\rm AV}=1.0$ is not suitable for model H24. Fig.~\ref{sedovr-fig} also shows model H24, but with $\beta^{\rm AV}=1.0$. Although this model roughly reproduces the density profile of the analytic solution; the density and pressure show significant scatter, especially in the high-resolution run. Since in galaxy simulations we include radiative cooling which is sensitive to the density, we conclude that this model is unacceptable for our purpose. Finally, the bottom panels of the figure shows that if the timestep limiter suggested by \citet{sm09} is not adopted, the code gives an incorrect density and pressure profile, as also demonstrated by \citet{sm09} and \citet{dd12}. This is because the particles in the cold and homogeneous interstellar medium are allowed to integrate their hydrodynamics equations with a larger timestep. The expanding shells can pass these particles before their subsequent integration time occurs. This will lead a massive underestimate of the effect of feedback in galaxy simulations. We stress that the individual timestep limiter must be implemented within SPH codes for galaxy simulations where the strong feedback from stellar wind and supernovae are included.

%\clearpage

%%% KHI plot

\begin{figure}
\centering
\includegraphics[width=0.8\hsize]{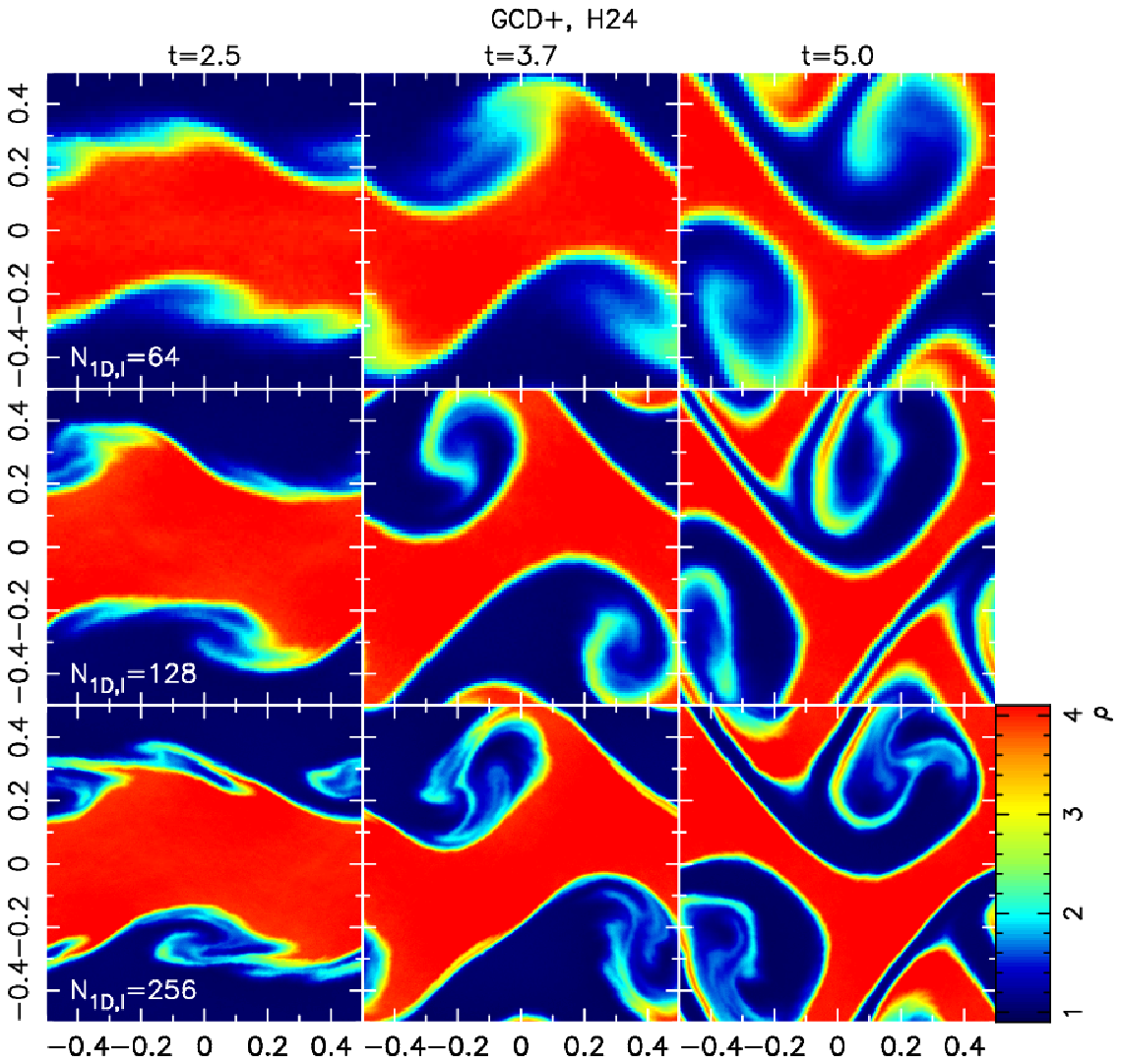}
\includegraphics[width=0.8\hsize]{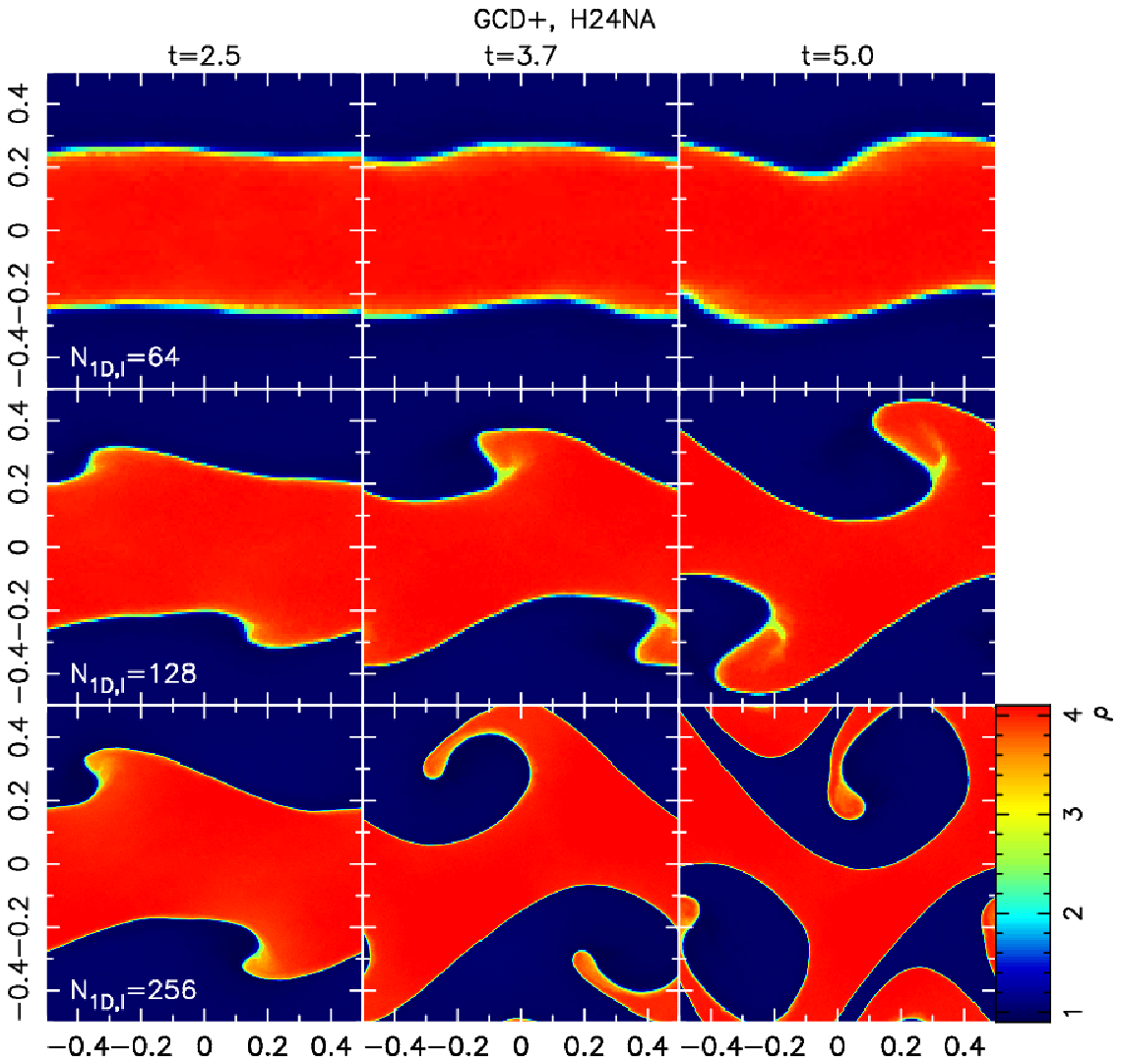}\\
\includegraphics[width=0.8\hsize]{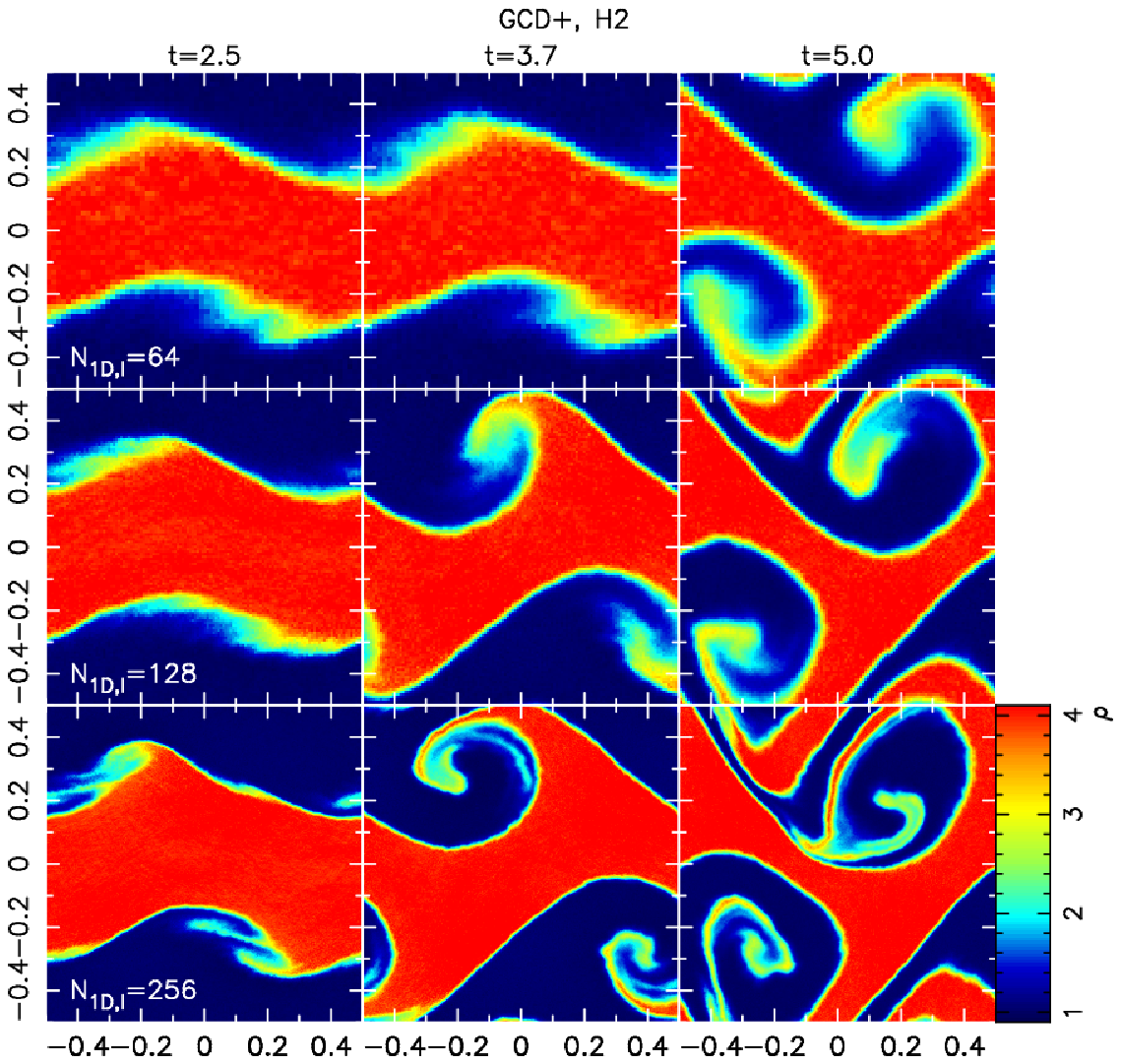}\\
\caption{
Density distributions at $t=t_{\rm KH}=2.5$ (left), $t=3.7$ (middle) and $t=5.0$ (right) with models H24 (top panels), H24NA (middle panels) and H2 (bottom panels). Top, middle and bottom panels in each panel show the results of simulations with different numbers of particles of $N_{\rm 1D,l}=64$, $128$ and $256$, respectively.
}
\label{wt-fig}
\end{figure}

\subsection{KHI test}

\citet{ams07} introduced a straightforward test which allows for a given
code (particle- or grid-based) to be assessed in terms of its ability
to resolve KHI \citep[see also][]{dp08,jwh10,vdrrd10,mlp12}. In this section,
we demonstrate that the updated {\tt GCD+} can resolve such instabilities. 
Following \citet{dp08}, we consider a two dimensional periodic boundary region
with $x=\{-0.5,0.5\}$ and $y=\{-0.5,0.5\}$.
The region within $|y|<0.25$ is set to be the high-density region
(with $\rho_{\rm h}=4$), while the rest is the low-density region (with
$\rho_{\rm l}=1$). Equal mass particles are adopted in both regions,
and $N_{\rm 1D,h}$ ($N_{\rm 1D,l}$) particles are used to cover the $x$-axis for the high-density
(low-density) region. The two regions are in pressure equilibrium and we assume $P_{\rm h}=P_{\rm l}=2.5$.
The high-density (low-density) region has velocity $V_{x,\rm h}=-0.5$ ($V_{x,\rm l}=0.5$).
We also added sinusoidal perturbations to the vertical velocity,
using $v_y(x)=\delta v_y \sin(\lambda 2 \pi x)$, setting
$\delta v_y=0.01$ and $\lambda=1.0$.
% These perturbations are only imposed to the particles within $|y|={0.225,0.275}$.
As before, we assume $\gamma=5/3$.
Following \citet{dp08}, we consider a time scale of KHI as
%%%%%%%%%%   equation
\begin{equation}
\tau_{\rm KHI} = 2 \pi/\omega
\label{tkhi-eq}
\end{equation}
where
%%%%%%%%%%   equation
\begin{equation}
 \omega=\frac{2 \pi}{\lambda}\frac{(\rho_{\rm h} \rho_{\rm l})^{1/2} | V_{x,\rm h}-V_{x,\rm l}|}
 {(\rho_{\rm h}+\rho_{\rm l})}.
\label{khiomg-eq}
\end{equation}
Our initial condition leads to a timescale for KHI of $\tau_{\rm KHI}=2.5$ and we run simulations for $t=2 \tau_{\rm KHI}=5.0$.
%  Notice that several authors including \citet{vs10b} use $\pi/\omega$ instead of $1/\omega$ in equation(\ref{tkhi-eq}).

 Since there is no analytic solution for this test. We compare the results of our code to those of a publicly available mesh code, {\tt Athena} \citep{sgths08}. In the {\tt Athena} runs, we chose the HLLC Riemann solver and third-order interpolation. We set the same initial condition as above for the {\tt Athena} runs. However, as discussed in \citet{rkgar10}, it is important for grid codes to initially resolve the contact discontinuity. Following \citet{vs10b} and \citet{rkgar10}, we apply the following 'ramp' function to the density and velocity
%%%%%%%%%%   equation
\begin{equation}
 R(y)=\frac{1}{1+\exp[2(y-0.25)/\delta_y]}\frac{1}{+\exp[2(y+0.25)/\delta_y]}.
\label{khiramp-eq}
\end{equation}
We run the two cases with $\delta_y=0.01$ and 0.05. In addition, to test Galilean invariance we also run the case where the whole region is moving with $V_{x,0}=100.0$.
\citet{vs10b} argues that applying the smooth change of density at the contact discontinuity is also important for the SPH simulations \citep[see also][]{vdrrd10}. However, it is difficult to assign such density profile in the SPH run without changing the particle masses, which we do not prefer to do because equation (\ref{h-eq}) is designed for the case that all the SPH particles have the same particle mass. Instead, we modify the thermal energy, and therefore entropy after calculating the initial density with the SPH kernel, so that the pressure is constant initially. This roughly corresponds to $\delta_y=0.01$ for our lowest resolution simulation case. 

\begin{figure*}
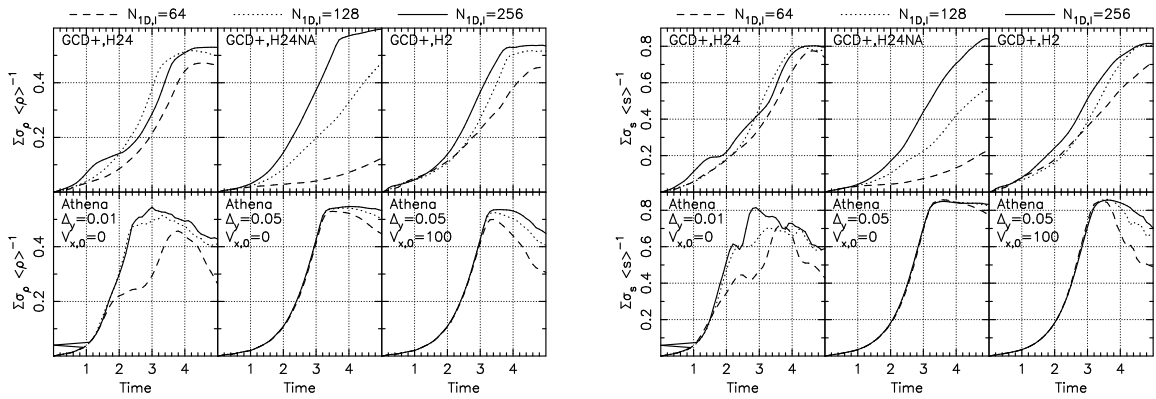

\centering
\includegraphics[width=0.45\hsize]{f12a.ps}
\includegraphics[width=0.45\hsize]{f12b.ps}
\caption{
Density (left) and entropy (right) mixing statistics (see text) as a function of time for different models. 
Top panels show the results of models H24, H24NA and H2. Bottom panels show the results of the {\tt Athena} code with $\Delta_y=0.01$ and $V_{x,0}=0.0$, $\Delta_y=0.05$ and $V_{x,0}=0.0$ and $\Delta_y=0.05$ and $V_{x,0}=100.0$
Dashed, dotted and solid lines show the results with number of particles of $N_{\rm 1D,l}=64$, $128$ and $256$, respectively.
}
\label{wtkhi2dmix-fig}
\end{figure*}

 To quantitatively compare the results, we calculate the mixing statistics for a property $f$, such as density, suggested by \citet{rkgar10} as follows. First, the average $<f>$ and dispersion $\sigma_f$ for each row are calculated. Then, the ratio of $\sigma_f/<f>$ are averaged by
%%%%%%%%%%   equation
 \begin{equation}
 \sum \sigma_f / <f> =\frac{\sum_i (\sigma_f/<f>)_{i} dy }{N_{y} dy},
\label{khisigf-eq}
\end{equation}
where $(\sigma_f/<f>)_{i}$ is $\sigma_f/<f>$ for row $i$, $dy=(L=1.0)/N_y$ is the grid size and $N_y$ is the number of grids along the $y$-axis. For {\tt GCD+} runs, we measure the property smoothed with the SPH scheme in the $N_{\rm 1D,l}\times N_{\rm 1D,l}$ grid. We calculate the mixing statistics for both density and
entropy, $s=P/\rho^\gamma$, following \citet{rkgar10}, and shown in Fig.~\ref{wtkhi2dmix-fig}.

\begin{figure*}
\centering
\includegraphics[width=0.4\hsize]{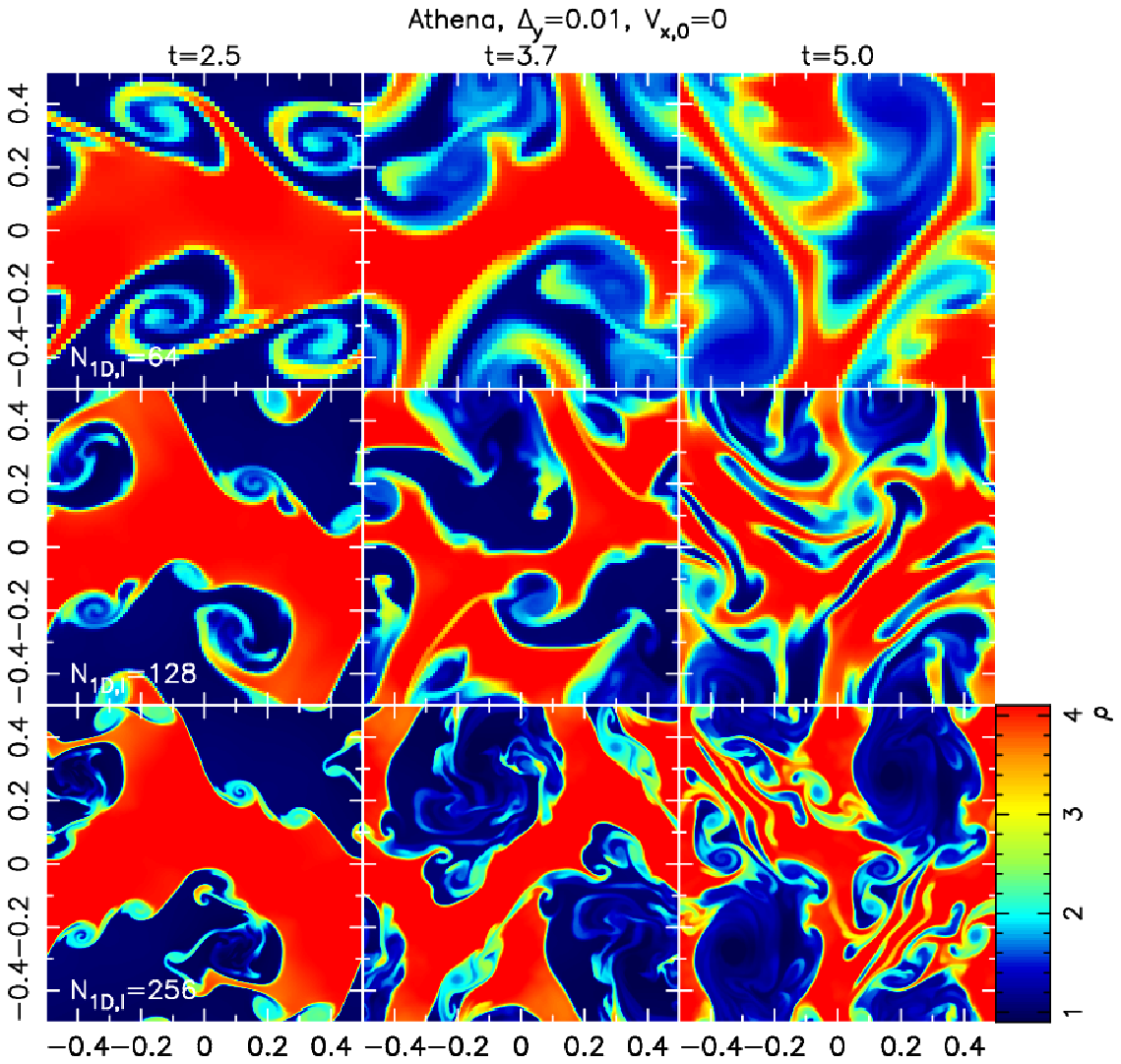}
\includegraphics[width=0.4\hsize]{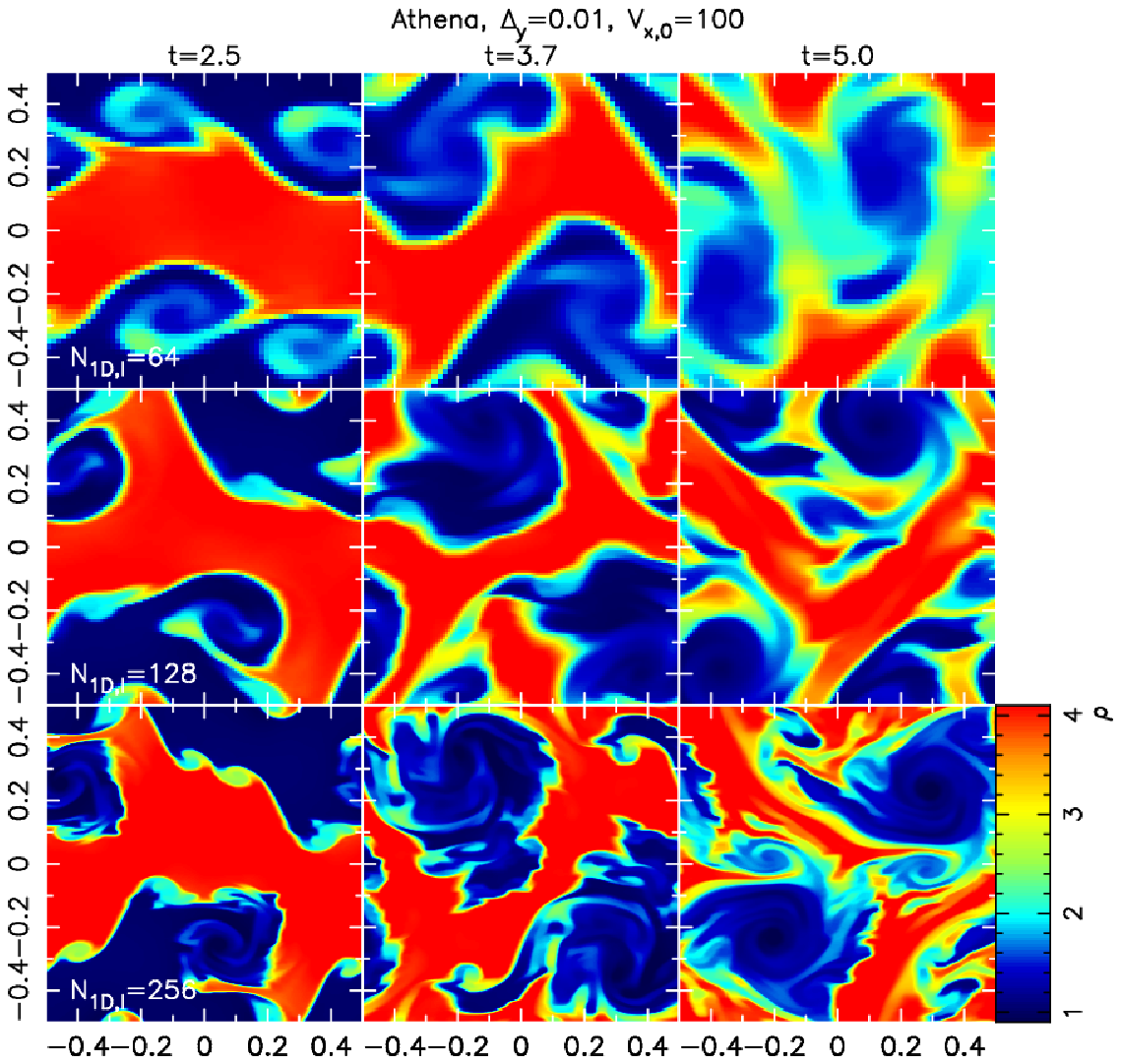}\\
\includegraphics[width=0.4\hsize]{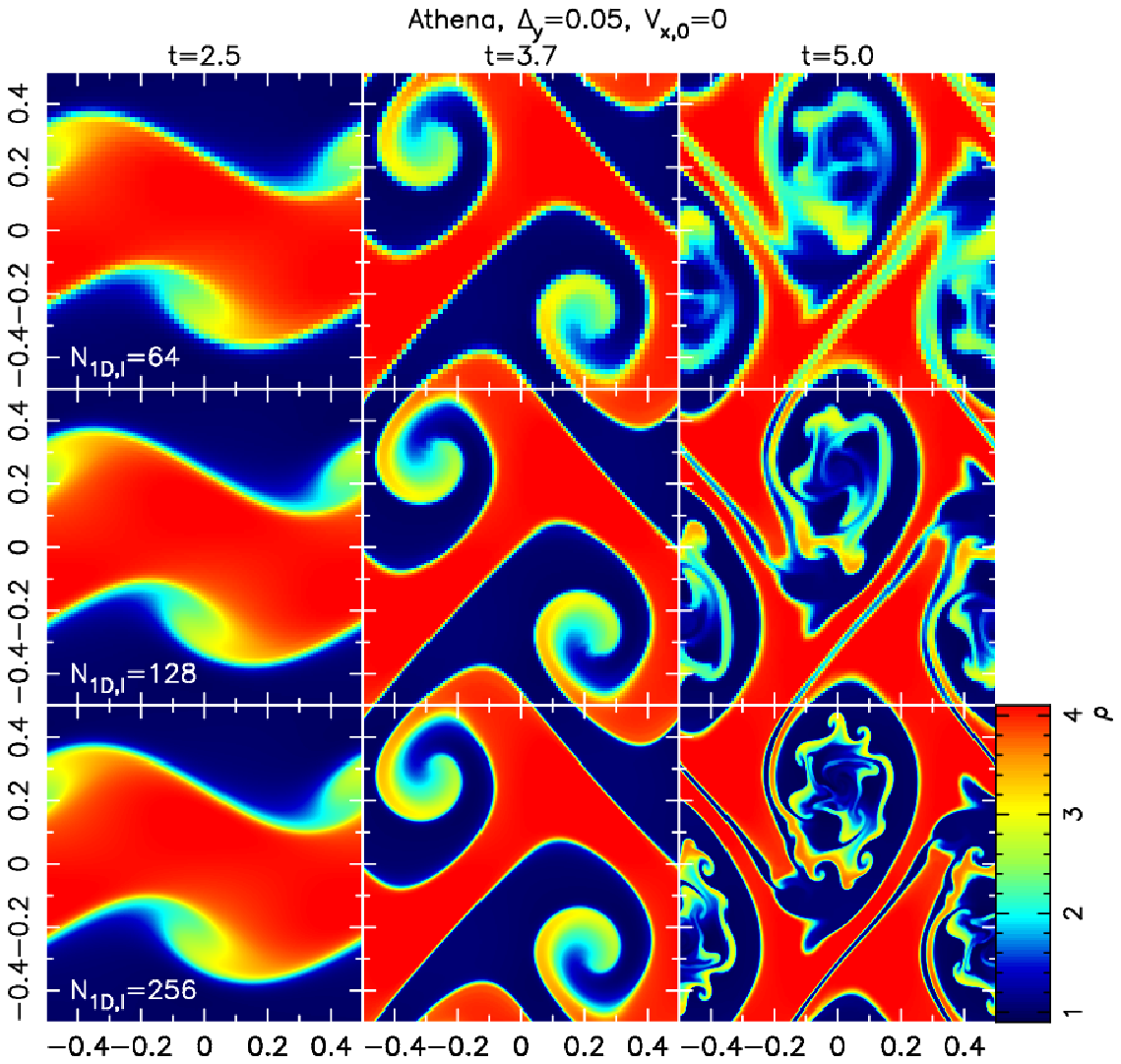}
\includegraphics[width=0.4\hsize]{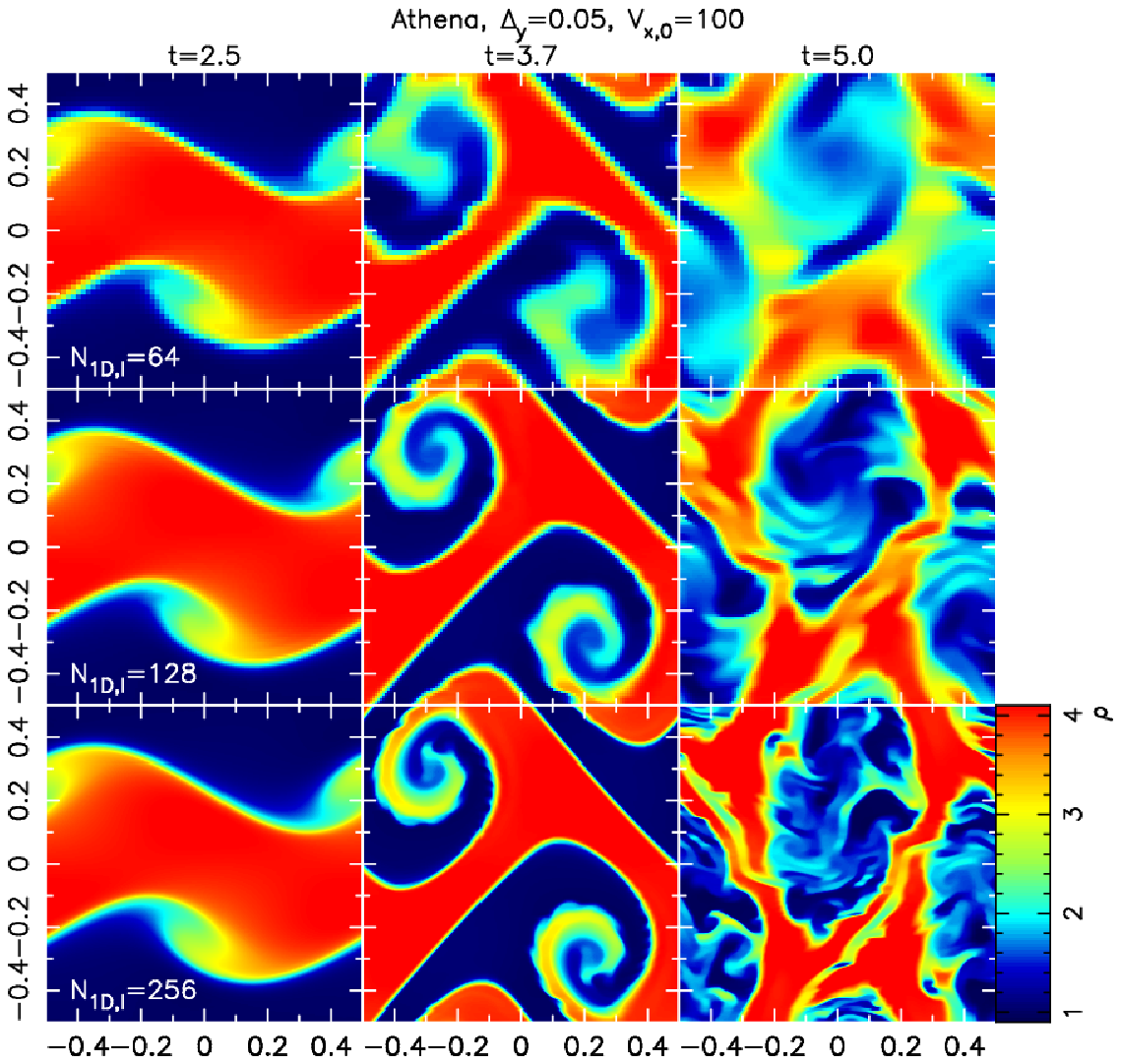}
\caption{
 As in Fig.~\ref{wt-fig}, but for the {\tt Athena} code results with $\Delta_y=0.01$ and $V_{x,0}=0.0$ (top left panels),
 $\Delta_y=0.01$ and $V_{x,0}=100.0$ (top right panels), $\Delta_y=0.05$ and $V_{x,0}=0.0$ (bottom left panels)
 and $\Delta_y=0.05$ and $V_{x,0}=100.0$ (bottom right panels).
Top, middle and bottom panels in each panel show the results of simulations with $N=64^2$, $128^2$ and $256^2$ grid, respectively.
}
\label{wta-fig}
\end{figure*}

 Fig.~\ref{wt-fig} demonstrates that the updated {\tt GCD+} is capable of capturing KHI,
and leads to similar results to those of the {\tt Athena}, such as shown in Fig.~\ref{wta-fig}, especially untill $t=3.7\sim1.5 t_{\rm KHI}$. At the later times, the grid code develops instabilities at smaller scales, said instability depends upon resolution (see Fig.~\ref{wtkhi2dmix-fig}). Fig.~\ref{wta-fig} shows that if the initial density profile was not smoothed enough, the small scale instability develops faster, and lead to the resolution dependent results. Fig.~\ref{wta-fig} also demonstrates that the development of grid size dependent small scale instabilities is sensitive to their global velocity field, i.e. Galilean non-invariance. However, if we apply enough smoothing to the initial density profile, i.e. $\delta_y=0.05$, the results are not sensitive to the velocity field or resolution, up to $t=3.7\sim1.5 t_{\rm KHI}$. Mixing statistics shown in Fig.~\ref{wtkhi2dmix-fig} demonstrate it quantitatively.

 In Fig.~\ref{wtkhi2dmix-fig}, model H24 shows a similar level of mixing as the {\tt Athena} results. There is a small dependence on the resolution, and we can see in  Fig.~\ref{wt-fig} that the small scale perturbations grow especially in the higher resolution run. 
 Model H24NA shows in Fig.~\ref{wt-fig} that without AC or the AV switch, normal SPH can still handle the rough features of KHI. However, as seen in the mixing statistics, it depends heavily on the resolution. Also, the features are much less smooth and the mixing of the two phases seems not to take place. Model H2 in Fig~\ref{wt-fig} demonstrates that even with a low $\eta$ model the KHI is captured with the new version of {\tt GCD+}.  However, compared to model H24, some resolution-dependent behaviour remains
 (Fig.~\ref{wtkhi2dmix-fig}); we can see that higher $\eta$ aids in the capture of KHI.

%%% Athena

%%% wtkhi2d mixs

\begin{figure}
\centering
\includegraphics[width=\hsize]{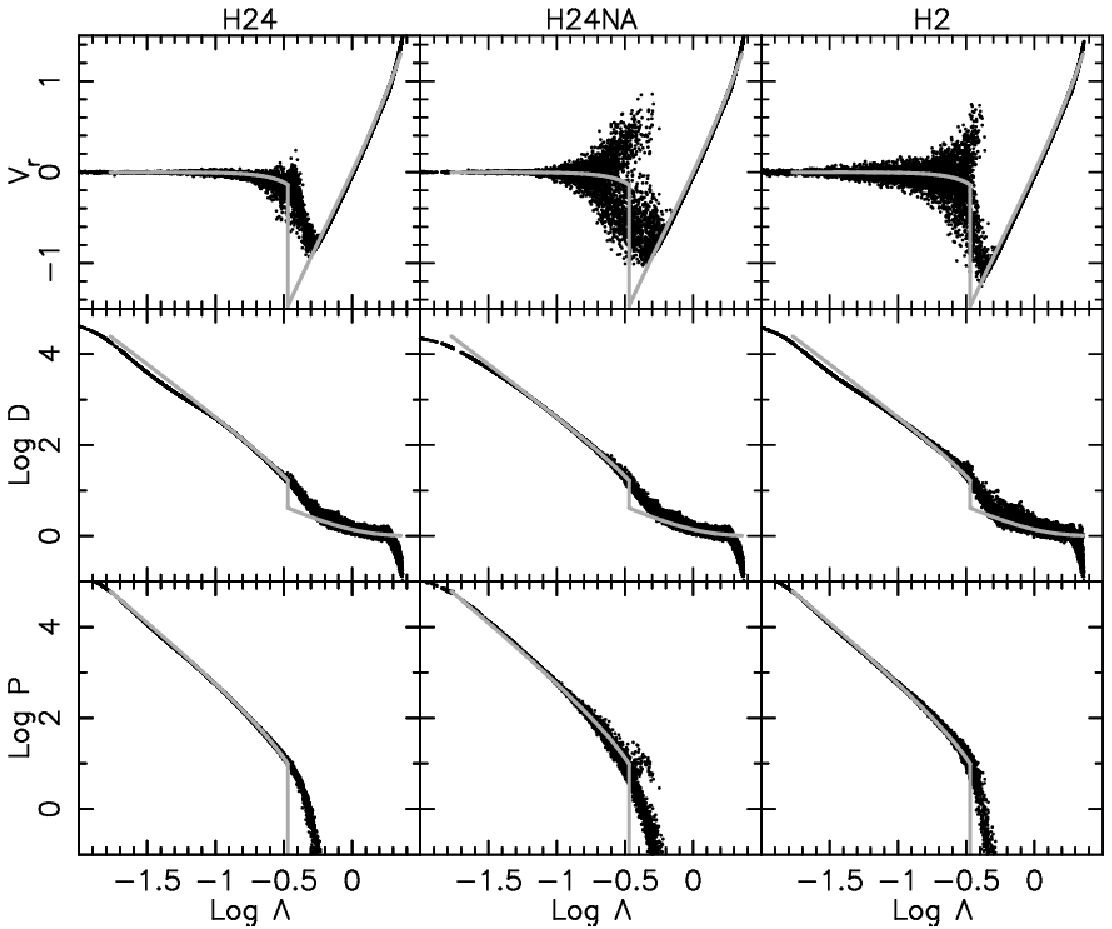}
\caption{
Normalised velocity (upper), density (middle), and pressure (lower)
distribution at an arbitrary time employing the Bertschinger (1985)
self-similar collapse test with models H24 (left), H24NA (middle) and H2 (right),
using $N=17,162$ particles.
The grey line represents the analytic solution.
}
\label{b85lam32-fig}
\end{figure}

\begin{figure}
\centering
\includegraphics[width=\hsize]{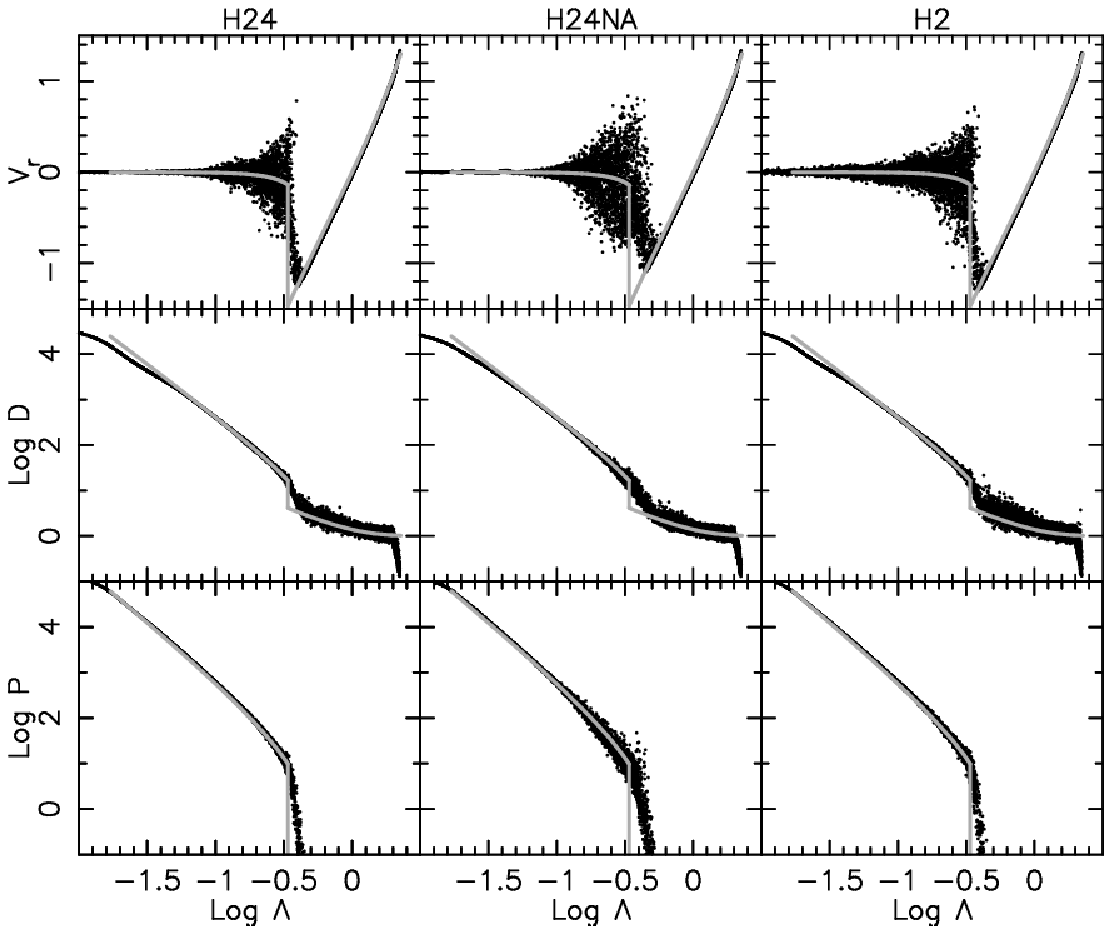}
\caption{
 Same as Fig.~\ref{b85lam32-fig}, but for the higher resolution simulations with $N=137,145$.
Only $1/8$ particles are shown to make a comparison with Fig.~\ref{b85lam32-fig} easy.
}
\label{b85lam64-fig}
\end{figure}

\subsection{Self-similar collapse test}
\label{ssc-sec}

To see the performance of the updated SPH scheme with self-gravity, we next run so-called self-similar collapse test. \citet{eb85} derived a self-similar solution for the collapse
of an overdense perturbation in an Einstein-de Sitter ($\Omega=1$) Universe.
\citet{nw93} introduced a test simulation based upon this self-similar
solution. Following \citet{nw93}, we consider a spherical volume
which initially follows the Hubble expansion, and 
set a central spherical perturbation with mass of $0.05 M_{\rm tot}$ and 
radius of $0.1 R_{\rm ini}$, where $M_{\rm tot}$ is the total mass
and $R_{\rm ini}$ is the initial radius of the simulation sphere.
To focus on testing hydrodynamics performance, we consider a pure gas collapse case, i.e. $\Omega_{\rm b}=1.0$, and no dark matter.
We set a glass-like distribution of the particles to describe the initial sphere with two different resolutions where employ $N=17,162$ and 137,145 particles respectively.

The dimensionless parameters for radius, $\Lambda$, radial velocity, 
$V_{\rm r}$, density, $D$, and pressure, $P$, are defined by
%%%%%%%%%%   equation
\begin{eqnarray}
 \Lambda (r,t) & = & \frac{r}{r_{\rm ta} (t)}, \nonumber \\
V_{\rm r}(\Lambda)& = & \frac{t}{r_{\rm ta}} v_{\rm r} (r,t)  , \nonumber \\
 D (\Lambda) & = & \frac{\rho (r,t)}{\rho_{\rm H}}, \nonumber \\
 P(\Lambda)  & =& \left(\frac{t}{r_{\rm ta}}\right)^2 
 \frac{p(r,t)}{\rho_{\rm H} }.
\label{ssv85-eq}
\end{eqnarray}
\noindent
Figs.~\ref{b85lam32-fig} and \ref{b85lam64-fig} show the results in these dimensionless parameters at and arbitrary time when 3,570 and 30,843 particles are within the shock radius, $r_{\rm shock}=0.34 r_{\rm ta}$, for the lower and higher resolution simulations respectively in model H24.
Dots within each panel represent the simulation results for the gas particles,
while the grey lines correspond to the analytic solution of \citet{eb85}.
Figures demonstrate that although all the models reproduce the analytic solution, the radial velocity has
too much scatter around the shock front in models H24NA. Also model H24NA shows significantly larger scatter in pressure around the shock front. This is similar to the results seen in Section \ref{ple-sec}. It is interesting to note that AC significantly stabilises the oscillation in velocity and pressure around the shock. Although model H24 applies a relatively high $\alpha_{\rm AV,min}$, a comparison between models H24 and H24NA presents the benefit of the AV switch, and model H24 shows a sharper shock feature, especially visible in the high-resolution simulations. Model H2 shows an even sharper shock feature than model H24. However, the scatter in radial velocity is significantly larger in model H2, compared to model H24. Therefore, we conclude that model H24 is superior to model H2, and model H24 is our best model. 

\subsection{Self-gravitating Gas Disc}
\label{isodisc-sec}

 Gresho vortex test in Section \ref{gresho-sec} shows a disappointing result. However, some basic test problems are often too critical. The target systems for our galactic science may not require the high-level of accuracy. In this section, we demonstrate that our best model, H24, achieves satisfactory angular momentum conservation in a disc galaxy simulation with self-gravity.
 
 We set up an isolated disc galaxy which consists of self-gravitating gas disc with no bulge component in a static dark matter halo potential. 
We use the standard Navarro-Frenk-White (NFW) dark matter halo density profile \citep{nfw97}, assuming  a standard cold dark matter ($\Lambda$CDM) cosmological model with cosmological parameters of $\Omega_0=0.266$, $\Omega_{\rm b}=0.044$ and $H_0=71{\rm kms^{-1}Mpc^{-1}}$, i.e. $h=0.71$:
\begin{equation}
\rho _{\rm DM}=(1-\Omega_b/\Omega_0)\frac{3H_{0}^{2}}{8\pi G}\frac{\rho _{c}}{cx(1+cx)^{2}},
\end{equation}
where
\begin{equation}
c=\frac{r_{200}}{r_{s}}, \;\; x=\frac{r}{r_{200}},
\end{equation}
and
\begin{equation}
r_{200}=1.63\times 10^{-2}\left(\frac{M_{200}}{h^{-1}{\rm M}_{\odot }}\right)^{\frac{1}{3}} h^{-1} \textup{kpc},
\end{equation}
where $\rho _{c}$ is the characteristic density of the profile, $r$ is the distance from the centre of the halo and $r_{s}$ is the scale radius. The halo mass is set to be $M_{200}=10^{12}M_{\odot }$ and the concentration parameter is set at $c=10$.

 The gaseous disc is set up following the method described in \citet{sdh05b}. 
The radial surface density profile is assumed to follow an exponential law with a scale length of $R_{d}=4$ kpc and the total gas mass of $10^{10}$ M$_{\odot }$. The initial vertical distribution of the gas is iteratively calculated to reach hydrostatic equilibrium assuming the constant temperature of $T=10^{5}$ K. We chose the relatively high temperature initially, to generate a stable gas disc and avoid non-axisymmetric structures to develop. We run the simulations with different numbers of particles, $N=10^{4}, 10^{5}$ and $10^{6}$.

\begin{figure}
\centering
\includegraphics[width=\hsize]{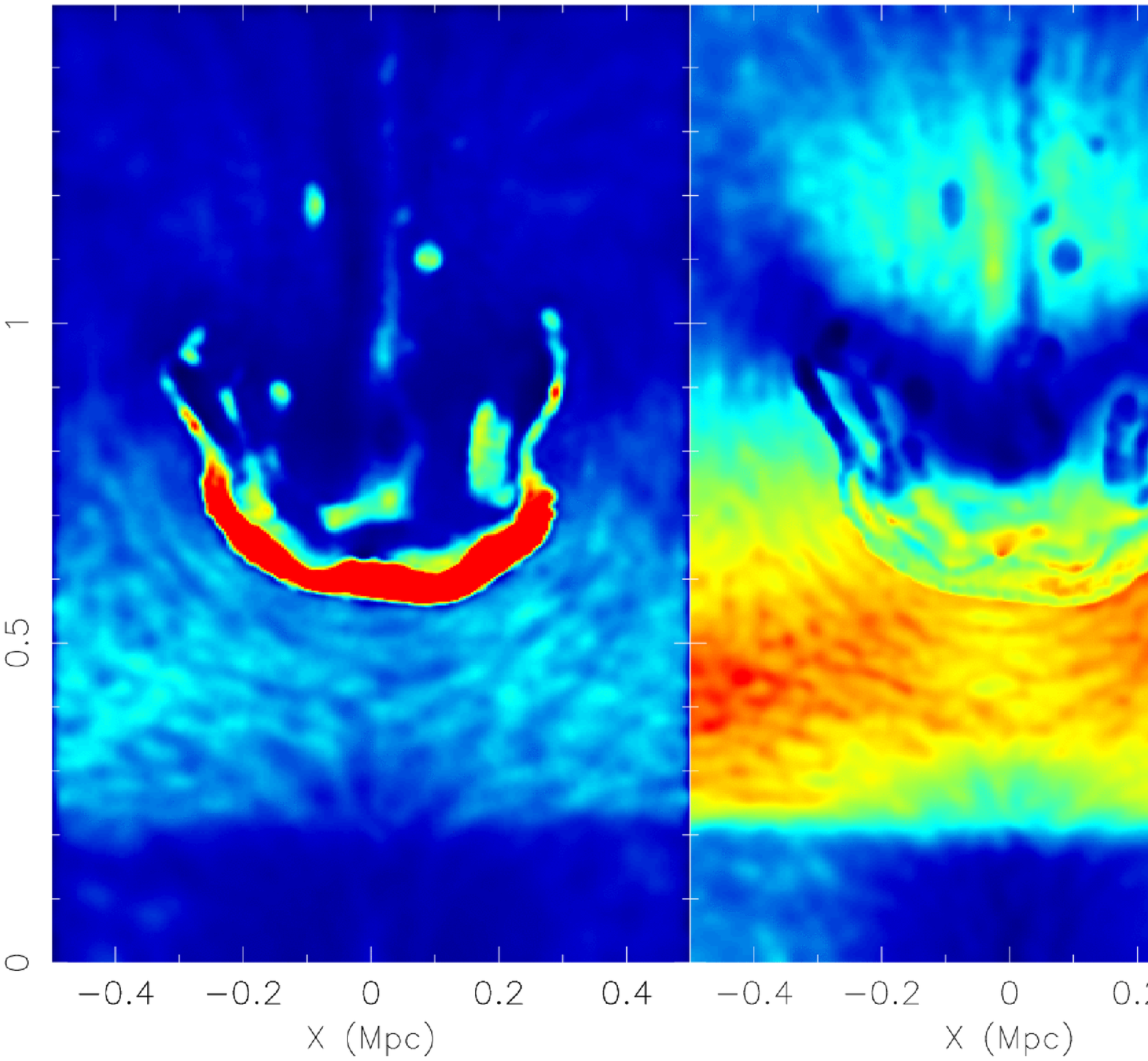}
\caption{
 Time evolution of the ratio, $R_{\rm JM,0.5}$ ($R_{\rm JM,0.25}$) between the radii that contains its half (quarter of) total mass and angular momentum. $R_{\rm JM,0.5}$ and $R_{\rm JM,0.25}$ are normalised to the initial values, $R_{\rm JM,0.5}(t=0)$ and $R_{\rm JM,0.25}(t=0)$. Left, middle and right panels show the results of the simulations with $N=10^{4}$, $10^{5}$ and $10^{6}$. In the left panels, solid, dotted and dashed line present the results of model H24, H24 without equation (\ref{dWc-eq}) and H24 without equation (\ref{fi-eq}). Middle panels show only two model results, i.e. models H24 and H24 without equation (\ref{dWc-eq}). Right panels show show only model H24 results. 
}
\label{idrjmt-fig}
\end{figure}

 Because the initial condition is not perfectly equilibrium, we run simulations for 1 Gyr, and let the system to relax. Then, using the relaxed system as an initial condition, we run the system for 2 Gyr. 
 This test is a similar in spirit to what is shown in Appendix of \citet{ns97}. Following to \citet{ns97}, we analysed the half-mass radius of the disc and the radius that contains half its total angular momentum, and name the ratio between these radii $R_{\rm JM,0.5}$. We also define more critical indicator, $R_{\rm JM,0.25}$, which is the ratio between the radii that contain a quarter of its total mass and angular momentum.
This ratio is expected to decrease if the angular momentum is transferred outward which brings the gas inward. Fig.~\ref{idrjmt-fig} shows the time evolution of $R_{\rm JM,0.25}$ and $R_{\rm JM,0.5}$ in different models and different resolutions. Solid lines of Fig.~\ref{idrjmt-fig} demonstrate that model H24 shows less than 10 \% of change in $R_{\rm JM,0.5}$, and less than 20 \% change in $R_{\rm JM,0.25}$ even in the lowest resolution simulations. Dramatic improvement is seen in higher-resolution simulations. In recent years, more than $10^5$ gas particles are often used to simulate the evolution of the gas disc \citep[e.g.][]{gkc12b}, and then the numerical angular momentum transport is minimal. Dashed line in the left panels of Fig.~\ref{idrjmt-fig} demonstrate that a significant angular momentum transport is observed if the velocity shear-corrected AV of equation (\ref{fi-eq}) is not applied \citep[see also Appendix of][]{ns97}. Dotted lines in the left and middle panels of Fig.~\ref{idrjmt-fig} indicate that although it is a tiny difference, there is systematically more angular momentum transport, because of the pairing instability, if the constant kernel gradient in equation (\ref{dWc-eq}) is not adopted. {\bf As mentioned above,} in more general galaxy simulations, the minimum smoothing and softening are required to be applied, which lead to a large number of neighbour particles within a fixed smoothing length and enhance the pairing instability. Therefore, in practice we need equation (\ref{dWc-eq}). Again it seems promising that a more sophisticated kernel function \citep[e.g.][]{da12} can minimise angular momentum transport and paring instability without the constant kernel gradient. However such kernels should be tested also in more practical simulations with self-gravity and radiative cooling. We wish to explore this in a future study.

\section{Summary} 
\label{sum-sec}

 We implement a modern treatment of SPH into our galactic chemodynamics code, {\tt GCD+}, in particular,
new AV and AC.
In this paper, we study how these new schemes work within the context of hydrodynamics simulations, 
and focus on the effect of the combination of the AC, the AV switch and the size of smoothing length.

We demonstrate that the AC and the AV switch help to ``smooth'' the thermal energy at the contact discontinuity.
Because of this improvement, the new code succeeds in capturing KHI. 
In essence, this confirms that the AV and AC scheme proposed by \citet{rp07} and
\citet{dp08} remedies the fundamental problem of SPH outlined by \citet{ams07}.
% It is also shown that to properly simulate the KHI, resolution remains 
% critical,
% even for our new implementation. Therefore, one must remain vigilant when
% it comes to resolution required in situations where
% KHI may become important.
 We also find that to capture strong shocks, like that expected in supernova explosions for example,
 the individual timestep limiter suggested by \citet{sm09} is crucial. 
 
 From these basic tests, we conclude that both models H24 and H2 are acceptable. 
However, in this paper, the pros and cons of these two models are highlighted. 
Model H2 with $\eta=2.0$, i.e. smaller smoothing length, resolves the shock features more sharply. However, we found from Gresho vortex tests that model H2 is less stable compared to model H24 with $\eta=2.4$, i.e. larger smoothing length. Also, model H24 captures KHI better, and is more stable when a strong shock is involved as demonstrated in point-like explosion and self-similar collapse tests. 
 Therefore, we favour model H24.

 In a forthcoming paper, we will carry out more realistic simulations of galaxy formation and evolution, including self-gravity, radiative cooling, star formation, SNe feedback and chemical
evolution, comparing and contrasting the behaviour of the different model parameters.

\section*{Acknowledgments}

The authors thank the anonymous referee for their valuable comments that improved the manuscript.
DK, DJB and BKG acknowledge the support of the UK's Science \& Technology Facilities Council (STFC Grand ST/H00260X/1, ST/F002432/1 and ST/H00260X/1). TO acknowledges the support by MEXT HPCI STRATEGIC PROGRAM and by Grant-in-Aid for Young Scientists (B) by JSPS (24740112). BKG acknowledges the generous visitor support provided by Monash University.  The calculations for this paper were performed on the Cray XT4 at Center for Computational Astrophysics (CfCA) of National Astronomical Observatory of Japan and the DiRAC Facility jointly funded by STFC and the Large Facilities Capital Fund of BIS. The authors acknowledge support of the STFC funded Miracle Consortium (part of the DiRAC facility) in providing access to the UCL Legion High Performance Computing Facility. The authors additionally acknowledge the support of UCL's Research Computing team with the use of the Legion facility. We thank James M. Stone for making the {\tt Athena} code publicly available. We also thank Eleuterio F. Toro and Kensuke Yokoi for providing the code to solve the Riemann problems.

\bibliographystyle{mn}
%\bibliography{../dkref}

\label{lastpage}

\end{document}